\documentclass[sigconf]{acmart}

\makeatletter
\def\@ACM@checkaffil{
    \if@ACM@instpresent\else
    \ClassWarningNoLine{\@classname}{No institution present for an affiliation}%
    \fi
    \if@ACM@citypresent\else
    \ClassWarningNoLine{\@classname}{No city present for an affiliation}%
    \fi
    \if@ACM@countrypresent\else
        \ClassWarningNoLine{\@classname}{No country present for an affiliation}%
    \fi
}

\usepackage{multirow}
\usepackage{makecell}
\usepackage{graphicx}
\usepackage{enumitem}
\usepackage[raggedrightboxes]{ragged2e}
\usepackage{tabularx}
\usepackage{tcolorbox}
\usepackage{algorithm}
\usepackage{algorithmic}
\usepackage{amsmath,amsfonts}
\usepackage{array}
\usepackage{caption}
\usepackage{subcaption} %
\usepackage{textcomp}
\usepackage{stfloats}
\usepackage{url}
\usepackage{verbatim}
\usepackage{hyperref}
\usepackage{siunitx}
\usepackage{booktabs}
\usepackage{balance}
\usepackage{xcolor}
\usepackage{colortbl}
\usepackage[utf8]{inputenc} 
\DeclareSIUnit{\coeq}{CO_2e}
\DeclareSIUnit{\metric}{metric}
\DeclareSIUnit{\tons}{tons}

\usepackage{soul}

\setcopyright{none}

\definecolor{color-A}{RGB}{0,0,0}
\definecolor{color-B}{RGB}{0,0,0}
\definecolor{color-C}{RGB}{0,0,0}
\definecolor{color-D}{RGB}{0,0,0}
\definecolor{color-E}{RGB}{0,0,0}
\definecolor{color-typo}{RGB}{0,0,0}
\definecolor{color-general}{RGB}{0,0,0}

\DeclareSIUnit{\coeq}{CO_2e}
\DeclareSIUnit{\metric}{metric}
\DeclareSIUnit{\tons}{tons}

\AtBeginDocument{%
  \providecommand\BibTeX{{%
    \normalfont B\kern-0.5em{\scshape i\kern-0.25em b}\kern-0.8em\TeX}}}

\begin{document}

\title{EcoServe: Designing Carbon-Aware AI Inference Systems}
\author{Yueying Li, Zhanqiu Hu, Esha Choukse\textsuperscript{\textdagger}, Rodrigo Fonseca\textsuperscript{\textdagger}, G. Edward Suh\textsuperscript{*}, Udit Gupta \\ \textit{Cornell Tech} \quad \textsuperscript{\textdagger}\textit{Microsoft Azure Research - Systems} \quad  \textsuperscript{*}\textit{Cornell and NVIDIA Research}}

\vspace{-0.1em}

\begin{abstract}
The rapid increase in LLM ubiquity and scale levies unprecedented demands on computing infrastructure. These demands not only incur large compute and memory resources but also significant energy, yielding large operational and embodied carbon emissions. In this work, we present three main observations based on modeling and traces from the production deployment of two Generative AI services in a major cloud service provider. First, while GPUs dominate operational carbon, host processing systems (e.g., CPUs, memory, storage) dominate embodied carbon. Second, offline, batch inference accounts for a significant portion (up to 55\%) of serving capacity. Third, there are different levels of heterogeneity across hardware and workloads for LLM inference. Based on these observations, we design EcoServe, a carbon-aware resource provision and scheduling framework for LLM serving systems.
It is based on four principles - \textbf{Reduce, Reuse, Rightsize, and Recycle (4R)}. With a cross-stack ILP formulation and design, we demonstrate that EcoServe can lower carbon emissions by up to 47\%, compared to performance, energy, and cost-optimized design points, while maintaining performance targets and SLOs. 

\end{abstract}
\maketitle

\pagestyle{plain}

\section{Introduction}

The rapid growth of Large Language Models (LLMs) has led to unprecedented demands on computing infrastructure.
From 2019 to 2022, we have seen an over 100$\times$ increase in model size from GPT-2 (1.5 billion parameters) to GPT-3 (175 billion parameters) to PaLM (540 billion parameters)~\cite{chowdhery2023palm}.
The increasing compute demands have also led to dramatic increases in the power and energy demands for deploying LLM's at-scale~\cite{tschand2024mlperf, luccioni2024power}.
The Electric Power Research Institute (EPRI) projects that AI-driven data centers will account for up to 40\% of the power in local grids (i.e. Virgina, Oregon)~\cite{epri2024}.

The rising compute demand of deploying LLMs impacts not only the efficiency of the data center but also the overall environmental impact of GenerativeAI and data centers.

\begin{figure}[t]
    \hspace{-0.2in}
         \centering
 \includegraphics[width=0.9\linewidth]{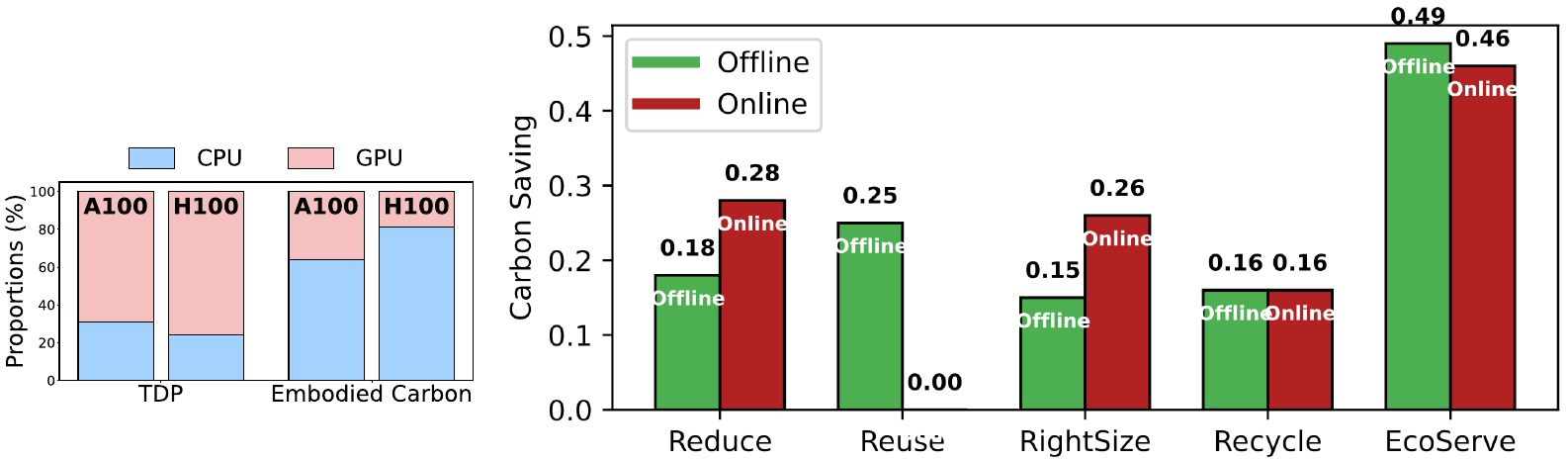}
 \vspace{-1em}
     \caption{(Left) Breakdown of thermal design power provision (TDP) and embodied carbon between host systems (CPU) and GPU. (Right) Through the 4R strategy, EcoServe optimizes the carbon savings for various AI workloads.}
     \label{fig:motivation}
     \vspace{-1em} 
 \end{figure}

The systems community has responded with numerous innovations to optimize LLM inference. Systems like vLLM~\cite{vllm2024perfupdate}, SGLang~\cite{zheng2023efficiently}, and DynamoLLM~\cite{stojkovic2024dynamollm} have significantly improved GPU utilization, throughput, and energy efficiency via sophisticated resource management. However, these approaches optimize primarily for operational metrics, overlooking a crucial aspect of AI's environmental impact: embodied carbon emissions from hardware manufacturing.

Recent sustainability reports from hyperscalers reveal that embodied carbon now accounts for over 50\% of their total emissions when using renewable energy~\cite{gupta2021chasing,microsoftSustainabilityReport,GoogleSustainabilityReport}. This presents a fundamental challenge: optimizing operational efficiency alone may not minimize total carbon impact. Our analysis reveals a critical insight: while GPUs dominate operational power consumption, CPUs and host systems contribute disproportionately to embodied carbon (Figure~\ref{fig:motivation}). Note that TDP can serve as
a proxy for operational carbon assuming a similar high utilization workload profile and same carbon intensity (assuming 80\% for both CPU and GPU).
The high embodied overheads from host systems owe to a combination of memory capacity, storage, and mother-board (e.g., PCB, peripheral interconnects). This misalignment creates an opportunity to rethink how we design AI infrastructure.

 Furthermore, AI inference workloads exhibit workload and hardware \textit{heterogeneity}—differences in model size, prompt length, decoding patterns, workload phases and SLO requirements—that create new optimization opportunities. %
 However, existing approaches fail to leverage all the heterogeneity dimensions, and co-optimize \textit{capacity planning, resource allocation, and runtime scheduling}, leading to misalignment between provisioned resources and runtime carbon reduction. For example, a strategy that only optimizes operational efficiency may increase embodied carbon by requiring frequent hardware upgrades or over-provisioning GPU capacity.

To holistically optimize embodied and operational emissions, we introduce \textbf{EcoServe}, a carbon-aware AI inference framework that co-designs \textit{capacity planning, resource allocation, and scheduling}. Unlike prior work that treats these phases independently, EcoServe’s \textit{cross-layer design} ensures that capacity planning decisions are effectively exploited at runtime. Specifically, while capacity planning determines long-term provisioning to reduce embodied carbon, runtime scheduling dynamically adapts these provisions to workload variation, ensuring optimal hardware utilization without increasing embodied or operational overhead.

At the core of EcoServe is a sustainability-driven optimization framework based on four key principles: \textbf{Reuse}, \textbf{Rightsize}, \textbf{Reduce}, and \textbf{Recycle} (4R). These principles collectively minimize carbon impact across different dimensions of system operation:
\begin{itemize}
    \item \textbf{Reuse}: Exploits underutilized host CPUs for offline inference to amortize embodied carbon across workload phases.
    \item \textbf{Rightsize}: Dynamically provisions GPUs and CPUs based on model size, execution phase, and workload demand to avoid over-provisioning.
    \item \textbf{Reduce}: Eliminates unnecessary host memory and storage resources to reduce embodied carbon overhead.
    \item \textbf{Recycle}: Extends the lifetime of host processing systems while selectively upgrading accelerators to balance embodied and operational trade-offs.
\end{itemize}

Figure~\ref{fig:motivation}(right) shows the impact of each of these design strategies on two scenarios: offline or batch processing, and online serving.
The individual strategies achieve carbon savings of 29-41\% compared to performance-optimized baselines.
While reduce and right size benefit carbon savings in online serving, offline serving benefits from a combination of reuse, reduce, and rightsize.
Altogether the different design strategies save up to 2$\times$ total carbon for both offline and online serving.
More generally, we evaluate EcoServe using both open-source datasets (ShareGPT~\cite{sharegpt}, Azure Functions~\cite{patel2023splitwise}) and production traces from a major cloud provider, as well as across heterogeneous model and hardware setups.
Our results demonstrate EcoServe's co-design solutions yield 1.4-2.2$\times$ reduction in total carbon emissions while maintaining performance targets across diverse LLM deployment scenarios.

This paper makes the following contributions:
\begin{itemize} [leftmargin=*]
\item We introduce EcoServe, a carbon-aware framework for LLM infrastructure design that optimizes both operational and embodied emissions. Fundamentally EcoServe comprises four design strategies---\textbf{Reuse, Rightsize, Reduce, and Recycle}---motivated by insights from fine-grained embodied carbon models for AI systems and detailed profiling and analysis of inference performance and efficiency across diverse hardware and serving environments.
\item Using EcoServe we show the individual design strategies save around 29\%, 25\%, 34\%, and 41\% carbon for reuse, rightsize, reduce, and recycle respectively, on a collection of open-source LLMs on heterogeneous hardware platforms, while maintaining SLO compared to a performance-optimized designs.
\item Putting the EcoServe optimizations together using an ILP formulation to co-optimize performance, efficiency, and carbon, we demonstrate 1.4-2.2 $\times$ carbon benefits with minimal performance degradation for both online and offline workloads across a variety of LLM models and workloads.
\end{itemize}

By framing AI inference sustainability as a \textit{co-design problem} across system stacks, EcoServe represents a paradigm shift towards carbon-aware systems design. Our results show that \textit{carbon-efficient AI infrastructure need not come at the cost of performance}—by leveraging workload and hardware heterogeneity, sustainable AI inference can be both high-quality and performant.

\section{Background}

\begin{table}[t]
    \centering \footnotesize
\begin{tabular}{ccc}\toprule
\textbf{Component} &\textbf{ Embodied kgCO$_2$e } & \textbf{Source(s)} \\ 
\midrule
SoC's & Tech. \& Area dependent & ACT~\cite{ACT}, iMec~\cite{imec-dtco20} \\ 
DDR4/LPDDR5 & 0.29 / GB& TechInsights~\cite{TechInsights} \\ 
GDDR6 & 0.36 / GB& TechInsights~\cite{TechInsights} \\ 
HBM2 & 0.28 / GB & TechInsights~\cite{TechInsights} \\ 
HBM3e & 0.24 / GB & TechInsights~\cite{TechInsights} \\ 
SSD & 0.110 / GB & Dell R740 LCA~\cite{dellr740} + SCARIF~\cite{SCARIF}\\ 
PCB & 0.048 / cm$^2$ 12 layers & Dell R740 LCA~\cite{dellr740}\\ 
Ethernet card &4.91 & Dell R740 LCA~\cite{dellr740} \\ 
HDD controller & 5.136 & Dell R740 LCA~\cite{dellr740} \\ 
Cooling & 7.877 per 100W & Scaled with TDP\cite{SCARIF} \\ 
PDN / PSU & 3.27 per 100W  & Scaled with TDP, Schneider~\cite{se_lifecycle_calculator} \\ \bottomrule
\end{tabular}
    \caption{Embodied carbon modeling strategy for each hardware component in AI inference systems. }
    \label{tab:model}
\vspace{-4em}
\end{table}

\begin{figure*}[t]
    \centering
\includegraphics[width=0.7\linewidth]{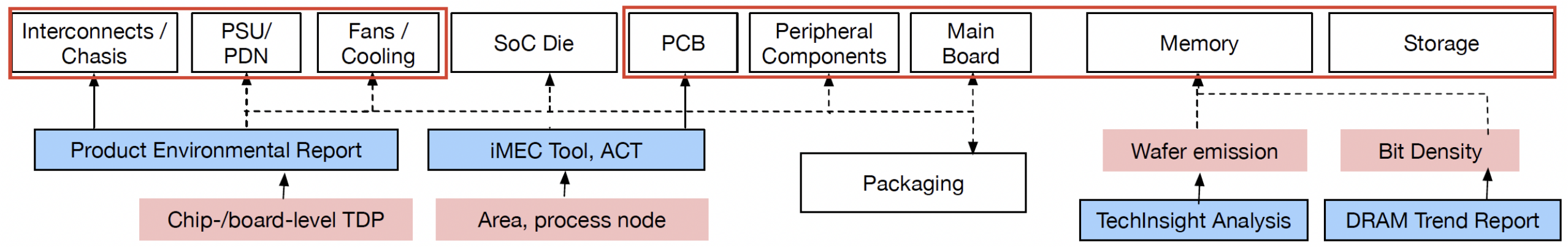}
    \caption{EcoServe's carbon modeling framework with more fine-grained embodied carbon estimation on memory, storage, and power-related components. We highlighted the differences with ACT in the red box.}
    \label{fig:framework}
\end{figure*}

\textbf{Carbon Emissions in Computing Infrastructure}. Computing infrastructure contributes to carbon emissions in two primary ways: operational carbon and embodied carbon. Operational carbon refers to emissions from energy consumption during system operation, including direct power consumption by computing components and indirect emissions from cooling systems. 

Embodied carbon, often overlooked, encompasses emissions from manufacturing, transportation, and eventual disposal of hardware components. Prior works indicate that for typical data center hardware, embodied carbon can represent over 50\% of total lifetime emissions, with memory and storage being particularly carbon-intensive in their manufacturing process~\cite{greensku, dirty_ssd, gupta2021chasing}.

Recent work like ACT, SCARIF, FOCAL, LLMCarbon~\cite{ACT,SCARIF,faiz2023llmcarbon,eeckhout2024focal} has highlighted the importance of considering both types of emissions. However, existing modeling approaches often do not look at fine-grained GPU carbon vs Host carbon footprints. We extend these works by providing a more fine-grained carbon modeling framework, especially for GPU systems, and show how these models can be used to optimize carbon footprint of LLMs    .

\textbf{LLM Serving Systems.}
State-of-the-art LLM serving systems like vLLM, TensorRT-LLM, AlpaServe, SGLang, Sarathi-serve, DistServe, and DeepSpeed~\cite{Deepspeed,kwon2023efficient,agrawal2024taming,zhong2024distserve,patel2023splitwise,zheng2023efficiently,TensorRT,alpaserve} have made significant advances in optimizing resource utilization LLM inference.

LLM inference typically has two distinct phases and two types of service level objectives (SLO): compute-bound prompt computation phase and memory-bound decoding phase. 
\textit{Online inference} prioritizes low-latency responses for interactive tasks, while \textit{offline inference} focuses on high-throughput batch processing, usually having 24hr SLO, presenting different optimization requirements. Additionally, the large model sizes and substantial key-value (KV) cache storage make LLM workloads memory capacity-bounded, further complicating resource management.

To address these challenges, these systems leverage GPU hybrid parallelization strategies, combining data, tensor, and pipeline parallelism to optimize computation across GPUs~\cite{mei2024helix,li2022amp}. Advanced memory management techniques, such as activation checkpointing, memory offloading, and paged attention, help alleviate memory pressure and improve resource utilization. Scheduler optimizations, including dynamic batching and chunked prefill processing, and disaggregating prompting and decoding phases, ensure high throughput while meeting latency constraints, enabling these systems to efficiently handle diverse LLM inference workloads~\cite{patel2023splitwise,zhong2024distserve}.

While these systems primarily focus on performance metrics such as throughput and latency, recent works like DynamoLLM~\cite{stojkovic2024dynamollm} and Melange~\cite{griggs2024mlange} have brought attention to energy and cost efficiency of LLM inference. However, optimizing for energy consumption or cost does not  directly optimize carbon impact. Carbon optimization goes beyond energy efficiency and cost considerations, requiring a holistic understanding of factors such as the carbon intensity of the energy grid, the lifetime utilization of the host (including CPU, memory, and storage) and the accelerator system, resource provisioning decisions, and the long-term environmental impacts of the underlying infrastructure.

\section{Understanding AI Systems' Carbon Footprint}

\begin{figure}[t]
    \centering
\includegraphics[width=0.9\linewidth]{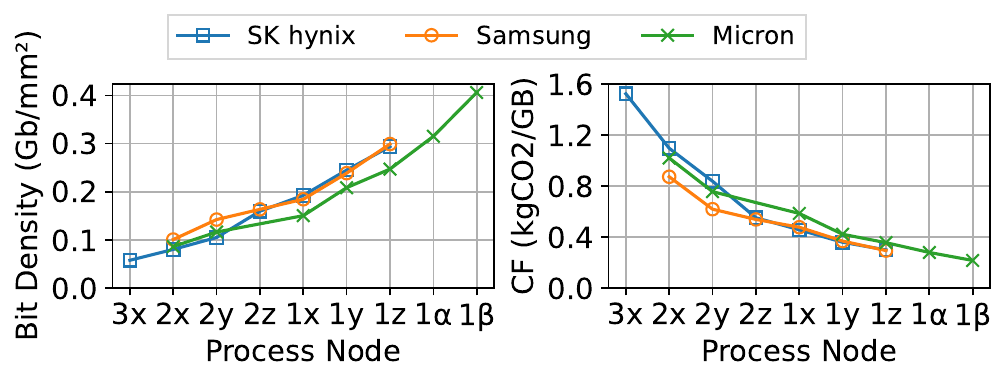}
\vspace{-1.5em}
    \caption{Trends in bit density (left) and embodied carbon footprint (right) across various DRAM memory technologies for 3 different manufacturers.}
    \label{fig:bitdensity_carbon}
    \vspace{-1em}
\end{figure}

Based on recent efforts quantifying the carbon footprint of computing systems~\cite{ACT, SCARIF, eeckhout2024focal}, the aggregate carbon emissions of AI systems is the combination of operational and embodied carbon as shown in the equation below:
{\footnotesize 
$$
    CF_{Task}  =  CF^{emb} + CF^{op} 
= (P_{host} + P_{gpu})\times t \times CI + CF^{emb}_{host} \frac{t}{LT} + CF^{emb}_{gpu} \frac{t}{LT}
 $$
}
where $P$ represents the power consumption of the host and GPU, $t$ represents the workload run-time, $LT$ represents the hardware lifetime, and $CF^{emb}$ represents the host and GPU embodied carbon.
In terms of operational emissions, we measure the power consumption of host-processing systems including CPUs, DRAM, SSD, using a combination of Intel RAPL~\cite{RAPL}; the power consumption of GPUs is empirically measured using NVML~\cite{nvml}.
The carbon intensity of the different data center power grids is based on geo-temporal variance of power grid carbon intensities~\cite{acun2023carbon, greensku, watttime,electricitymaps}.
To estimate embodied carbon, we leverage a combination of product environmental reports and life cycle analyses from industry~\cite{dellr740} and open-source tools to quantify the emissions from hardware manufacturing including the Architectural Carbon Modeling Tool (ACT), SCARIF, and GreenSKU~\cite{ACT, SCARIF, greensku}. There is a void in fine-grained modeling of different memory technology nodes for GPUs, which significantly impacts the embodied carbon footprint estimation. Furthermore, the peripheral components for cooling or power delivery are also not considered in the past models, which is essential for higher-end GPUs~\cite{ACT, SCARIF}.

In this section we describe how we extend current methods to study the embodied carbon of AI inference platforms including host systems, motherboards, and GPUs.

\subsection{Modeling Embodied Carbon of AI Systems}

The embodied carbon footprint (CF) owes to a combination of components for the host systems and GPUs. 
For host systems, we must account for the main CPU die, DRAM memory, SSD and HDD storage, printed wiring board (PWB), power delivery network (PDN), and server chassis.
Similarly, for the GPUs we must account for SoC, DRAM memory, PDN, cooling (e.g., heat sink), and PWB.
Table~\ref{tab:model} describes the salient parameters in the embodied carbon model for AI inference systems.
For many components including PDN and cooling (e.g., heat sink) given the lack of specific technology information we rely on publicly available product environmental reports from Dell~\cite{dellr740}.
Below we detail the methodology for estimating the embodied carbon of the remaining salient components. 

\subsubsection{Integrated circuits (Application Processors)} 
To compute the embodied carbon emissions of application processors (e.g., CPU die, GPU die) we rely on recently published architectural carbon modeling tools~\cite{ACT, SCARIF, imec-dtco20, eeckhout2024focal}.
For application processors we input the process technology and die area into ACT~\cite{ACT, imec-dtco20}.

\subsubsection{Memory and Storage}
Recent efforts from industry demonstrate memory and storage dominate the embodied carbon of Microsoft's data center hardware~\cite{greensku}.
To quantify the memory and storage embodied carbon, we consider both memory technologies and capacities.
For storage, we estimate SSD emissions based on the Dell R740 life cycle analysis~\cite{dellr740}, yielding an average estimate of 0.110 kg CO$_2$e per GB capacity; this is a conservative estimate compared to estimates of 0.160 kg CO$_2$e per GB from academic surveys~\cite{dirty_ssd}.
For memory, we extend existing ACT and SCARIF models by scaling wafer-scale DRAM data for different technologies from TechInsights~\cite{TechInsights,techinsights_micron_dram_2024,skhynixHynixNewsroom} by publicly available DRAM bit-densities per technology from memory vendors (e.g., SK Hynix, Micron, Samsung) as shown in Figure~\ref{fig:bitdensity_carbon}. 
The trends show newer technology nodes, given higher bit-densities, exhibit lower embodied carbon per GB capacity.
For the purposes of this work we estimate the DDR4/LPDDR5, GDDR6, and HBM2, HBM3e embodied carbon footprint as 0.29, 0.36, 0.28, 0.24 kg CO$_2$e per GB respectively.

\begin{figure}[t]
\includegraphics[width=0.9\linewidth]{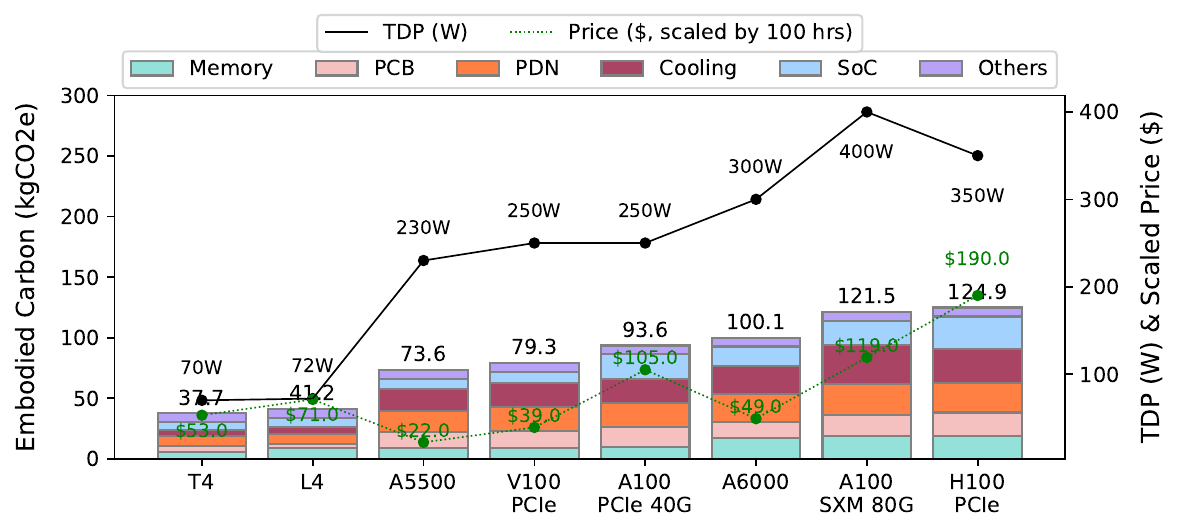}
\vspace{-1.5em}
    \caption{Trends in embodied carbon, power, and cloud cost for different generations of GPUs. As GPU performance increases (left to right), power consumption, cost and embodied carbon exhibit distinct trends, ACT only accounts for around 20\% in the blue SoC component ~\cite{aws-gpu,azure-gpu,lambdalabsCloudDeep}.  }
    \label{fig:cf_gpu}
    \vspace{-1em}
\end{figure}

\subsubsection{Mainboard Printed Wiring Board (PCB)}
To estimate the embodied emissions of the mainboard printed wiring board (PWB) we use the Dell R740 LCA~\cite{dellr740}. 
Given the PWB area of the Dell R740, 1925 square centimeters (176 kg CO$_2$), is significantly higher than modern server-class host systems and GPU's, we scale the PWB area for specific systems.
Thus, we estimate the embodied carbon of PWB to be:
$
    \text{Total PCB CF} = 0.048 \times Area_{PCB}
$.

\subsection{Characterizing Total Carbon}\label{sec:characterization}
Based on the carbon models and runtime energy measurement, we can study the carbon breakdown of AI inference platforms.

\textbf{Observation 1: Along with power capacity (TDP), embodied carbon of GPUs continues to rise as systems increase in compute and memory bandwidth. }

Figure~\ref{fig:cf_gpu}  analyzes the embodied carbon estimates of various NVIDIA GPUs, breaking it down into key components such as memory, printed circuit board (PCB), power delivery network (PDN), cooling systems, and system-on-chip (SoC). The breakdown aligns with findings from the recent TPU embodied carbon modeling paper~\cite{schneider2025life}.  Figure~\ref{fig:cf_gpu} illustrates these components using stacked bars to represent embodied carbon, with the thermal design power (TDP) plotted on the right axis. We observe a clear trend: with newer GPU generations, computational performance, memory bandwidth, and energy efficiency improve. However, the embodied carbon also rises with increasing computational performance, advanced thermal solutions required, and improved memory capacity. 
For instance, compared to an NVIDIA H100, and NVIDIA L4 incurs 3$\times$ lower embodied carbon (see Section~\ref{sec:rightsize}).

\textbf{Observation 2: Host systems levy high embodied carbon costs on end-to-end AI systems, primarily due to the mainboards, memory, and storage components.}

\begin{figure}[t]
    \centering
\includegraphics[width=0.9\linewidth]{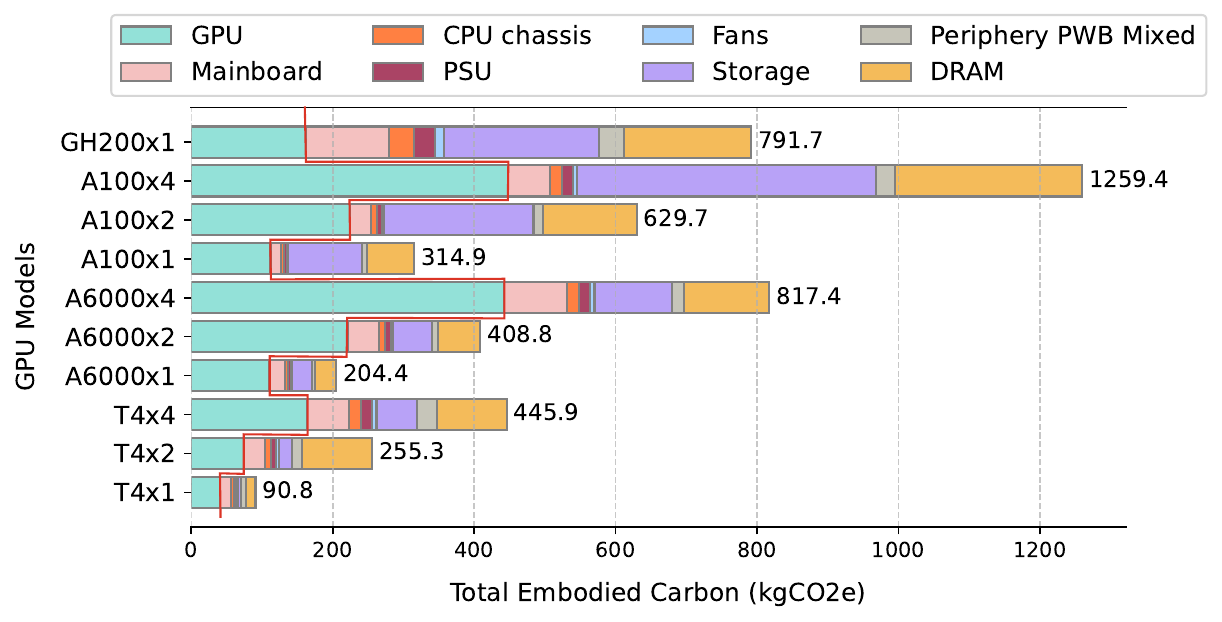}
\vspace{-1em}
    \caption{ Embodied carbon breakdown of full inference systems available in cloud offerings from Azure and LambdaLabs~\cite{azure-gpu,lambdalabsCloudDeep}, varying the number and type of GPU. Host-processing systems account for over half of the embodied carbon in AI systems, owing largely to memory, storage, and mainboard overheads.}
    \label{fig:cf_cpu_h}
    \vspace{-2em}
\end{figure}

Figure~\ref{fig:cf_cpu_h} shows the embodied carbon breakown of end-to-end inference systems including both GPUs and host systems.
The analysis highlights that the motherboard, DRAM, and storage constitute a significant portion of the embodied carbon impact.
However, for AI inference workloads, host systems are frequently underutilized compared to GPUs, leading to a disproportionate contribution of their embodied carbon footprint relative to their actual usage during operations.
This observation motivates exploring strategies such as resource sharing on the host (e.g., model multiplexing), improving utilization through CPU-based inference, extending the lifetime of host system components, and reducing the memory subsystem's footprint (Section~\ref{sec:reduce}).

\textbf{Observation 3: As low-carbon-intensity energy sources increasingly dominate hyperscalar data centers, the carbon footprint of AI infrastructure is increasingly determined from the embodied carbon of both GPUs and host systems.}

\begin{figure}[t]
    \includegraphics[width=0.8\linewidth]{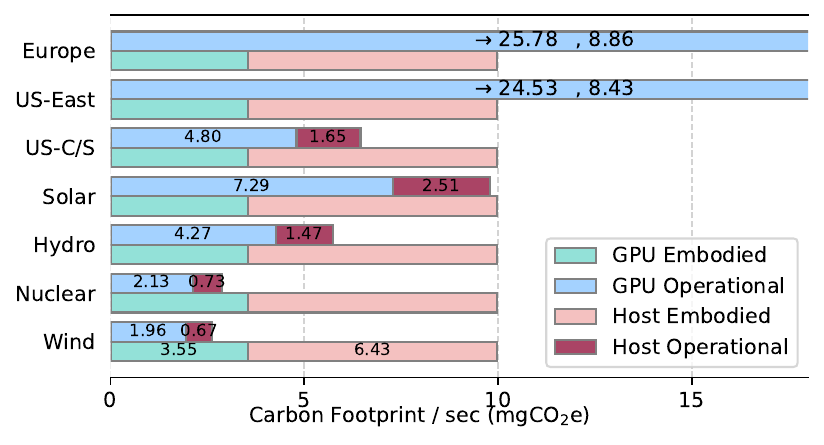}
    \vspace{-1.5em}
    \caption{Embodied and operational carbon breakdowns of an AI inference system comprising an NVIDIA A100 GPU. We assume a 4-year lifetime running Llama-13B. In data centers and power grids with high degrees of renewable energy (e.g., hyper-scalars), embodied carbon dominates the overall carbon footprint of AI systems.}
    \label{fig:cf}
    \vspace{-1em}
\end{figure}

Figure~\ref{fig:cf} compares embodied and operational carbon per second across datacenters with different power sources. The host and GPU embodied carbon remain fixed, while operational carbon varies with energy source carbon intensity. In high-carbon regions (Europe, US-East), GPU operational carbon dominates. As cleaner energy is adopted, operational emissions drop, making embodied carbon more significant. GPU power accounts for most operational carbon, while the host system dominates embodied carbon. Optimizing operational carbon requires improving GPU energy efficiency, minimizing idle power consumption, dynamic workload allocation, and clean energy integration. Reducing embodied carbon involves increasing host utilization, extending hardware lifetimes, and improving resource sharing to amortize the embodied carbon cost of host components (see Section~\ref{sec:recycle}).

\textbf{Observation 4: CPU is underutilized in AI inference systems.} During auto-regressive inference, CPUs are responsible for the following operations: (1) waiting for incoming requests and processing HTTP and GRPC operations, (2) input tokenization, (3) prefix matching and request ordering, (4) allocating tensors, memory, and metadata for batches, and (5) inference post-processing including de-tokenization, speculative decoding, and beam search processing.
Profiling several models (i.e., opt125m, Metallama-3-8B, Mistral and Mixtral) and request rates (2-32 requests per second from ShareGPT~\cite{sharegpt}) on vLLM (v0.6.2) and SGLang~\cite{zheng2023efficiently}, we find average CPU utilization is around 6\% of a 16-core processor (e.g., on average only a single CPU core is used).
Newer optimization like overlapped schedulers are proposed, so CPU processing is not on the critical path of recent LLM inference schedulers~\cite{vllm2024perfupdate}. For multimodality, encoder-based models or agentic systems, CPU can have a higher utilization and thus  in itself provide an opportunity to \textit{reuse the CPU side of the host system, while \textit{recycling} or \textit{reducing} the memory subsystem (See Section~\ref{sec:reuse}).} 

\section{EcoServe Design}~\label{sec:ecoserve}
\begin{figure*}[t]
\centering     \includegraphics[width=0.8\linewidth]{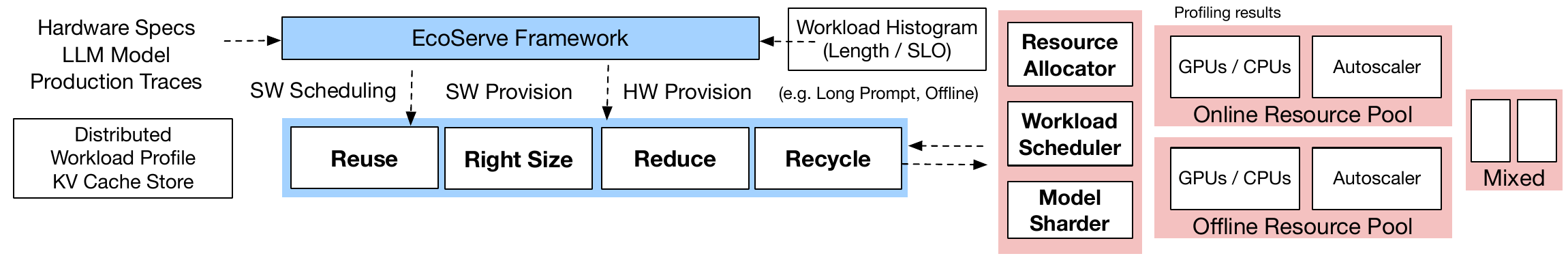} 
\vspace{-1em}
    \caption{EcoServe system diagram: An optimization framework that minimizes operational and embodied carbon across hardware resources through 4R co-designed strategies. Outputs inform scheduling and resource allocation decisions. }\label{fig:ecoserve}
    \vspace{-1em}
\end{figure*}
EcoServe is a carbon-aware framework for serving large language models (LLMs) that jointly optimizes performance, cost, and environmental impact through systematic resource management. The framework operates on four key principles: Reuse, Right Size, Reduce, and Recycle (4R), orchestrating both software and hardware provisioning decisions to minimize the carbon footprint of LLM inference. Figure~\ref{fig:ecoserve} shows the overall EcoServe design architecture.

EcoServe's framework ingests hardware specifications, LLM characteristics, and production traces alongside carbon intensity data to make carbon-aware scheduling decisions. The output such as number of GPUs, CPU cores for each workloads are directly usable for commodity resource scheduler or autoscaler. 

\subsection{Design Principles}
EcoServe implements four strategies for carbon-efficient inference:
\begin{itemize}[leftmargin=*]
\item \textbf{Reuse (\textit{Software Runtime}):} Leverages idle CPU capacity in existing AI inference systems for offline/batch workloads, increasing throughput while reducing resource demand and embodied carbon.
\item \textbf{Right Size (\textit{Software Provisioning}):} Optimizes GPU provisioning across heterogeneous hardware (L4, A6000, A100, H100, etc.) for both online and offline inference, matching resources to workload characteristics.
\item \textbf{Reduce (\textit{Hardware SKU Design}):} Minimizes waste by eliminating underutilized host compute, memory, and storage resources that contribute to high embodied carbon overhead.
\item \textbf{Recycle (\textit{Hardware Provisioning}):} Extends hardware lifetimes asymmetrically across host processing systems and GPUs to balance operational and embodied emissions.
\end{itemize}
These strategies provide a comprehensive solution across software runtime optimization, software provisioning, and hardware provisioning to reduce the carbon footprint of AI inference systems. Below we detail on each strategy, then present how the output of the strategy interacts with the components in real systems.
\subsubsection{\textbf{Reusing CPUs for Offline Inference}}~\label{sec:reuse}
Section~\ref{sec:characterization} shows that host processing systems incur high embodied carbon overheads in AI inference infrastructure.
To maximize the utility of these resources, EcoServe opportunistically reuses host CPUs for offline inference, reducing the reliance on high-power GPUs.

\textbf{Identifying opportunities for using CPUs.} 
Due to the underutilization of host resources, we must carefully determine which operations or inference phases can be offloaded from GPUs to CPUs to maximize carbon efficiency. Note that each offload incurs additional energy costs for data movement, necessitating a selective approach to CPU reuse. %

\begin{figure}[t]
    \centering
    \includegraphics[width=0.85\linewidth]{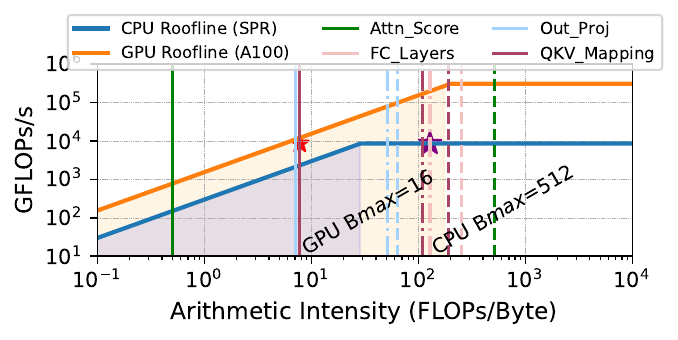}
    \vspace{-1em}
    \caption{ Roofline models of an Intel Sapphire Rapids CPU with 112 cores compared to an NVIDIA A100. 
    We overlay different operators in the decode (solid and dash-dot line) and prompt computation phase (dashed line) for Llama-3-8B with the maximum batch size that CPU/GPU can handle at a context length of 2048.
    At large batch sizes, the GPU is capacity bound ({\textcolor{red}{\(\star\)}}) while the CPU ({\textcolor{purple}{\(\star\)}}) can efficiently process offline decoding workloads.}
    \label{fig:roofline}
    \vspace{-2em}
\end{figure}

To guide this decision, Figure~\ref{fig:roofline} presents a {roofline model analysis~\cite{williams2009roofline,yuan2024llm,nerscRooflinePerformance}} comparing the computational intensity and memory bandwidth efficiency of an Intel Sapphire Rapids 112-core CPU and an NVIDIA A100 40GB GPU.
Given the relatively lower memory bandwidth gap between the CPU and GPU, we find that {low-arithmetic-intensity operations}, such as {attention scoring and KV cache decoding}, are well-suited for CPU execution.
Additionally, for large-batch operations, GPU throughput becomes capacity-bound, making CPU reuse {a practical alternative for offline inference.
Figure~\ref{fig:roofline} highlights the maximum batch sizes feasible for GPUs (16) and CPUs (512) at a context length of 2048 in FP16.}

\begin{figure}[!t]
    \centering
    \includegraphics[width=0.7\linewidth]{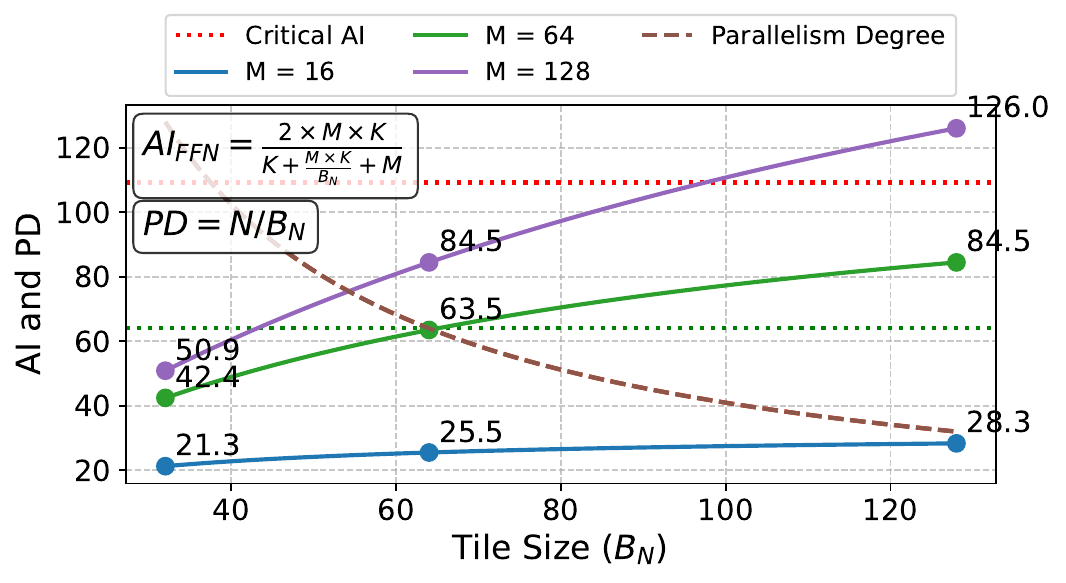} 
    \vspace{-1em}
    \caption{CPU offline inference requires carefully balancing parallelism degree (PD) and arithmetic intensity of Linear operators (AI, FLOPS/Byte). The run-time scheduler co-designs tile and workload slice dimensions with architecture decisions to optimize offline performance.}
    \label{fig:gemm-ai}
    \vspace{-1.2em}
\end{figure}

\textbf{Optimizing CPU performance via parallelization.}
To maximize the efficiency of CPU-based inference, EcoServe optimally configures thread-level parallelism and tensor tiling to balance:
\begin{itemize}
    \item Arithmetic intensity (FLOPs/Byte),
    \item Compute vs. memory bandwidth constraints,
    \item Model size, batch size, and sequence length.
\end{itemize}
The optimal configuration depends on the CPU microarchitecture (e.g., AMX and AVX efficiency), number format (FP6, BF6, INT8, etc), and workload characteristics.
As illustrated in Figure~\ref{fig:gemm-ai}, {parallelization along the KV sequence length dimension, in addition to the batch dimension, maximizes memory bandwidth utilization across all CPU cores.} 
This is particularly beneficial for long-context workloads, where attention mechanisms contribute significantly to inference latency~\cite{flashdecoding2023,flashinfer,dao2022flashattention,dao2023flashattention,hong2023flashdecoding++}.

Following our {roofline analysis}, while prompt computation remains GPU-favorable, EcoServe implements {layered pipelining} to move KV cache transfers from GPUs to CPUs for decoding.
By optimizing these parallelism dimensions, we achieve a {3.67$\times$ speedup} compared to a baseline Llama.cpp implementation~\cite{llama_cpp}, with an average {1.4$\times$ improvement} across model sizes and sequence lengths (Figure~\ref{fig:speedup-reuse}).
As CPU inference remains a nascent field, future advancements leveraging sparsity and weight-sharing techniques are expected to further enhance CPU-based offline inference~\cite{song2023powerinfer,zhang2024h2o,Liu2023DejaVu}.

\begin{figure}[h]
    \centering
    \includegraphics[width=0.95\linewidth]{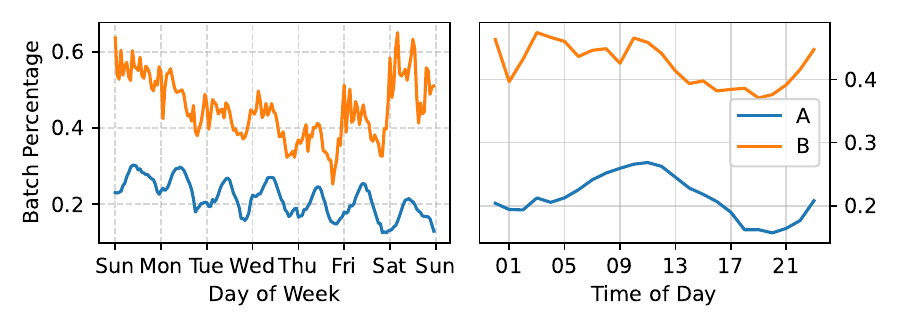}
    \vspace{-0.2in}
    \caption{Breakdown of online and offline demand for two LLM services (A and B) running in a production cloud data center over a week (left) and during a day (right). 
    Offline demand accounts for a significant portion of total serving capacity, creating opportunities for CPU reuse in offline inference.}
    \label{fig:offline}
    \vspace{-1.2em}
\end{figure}

\textbf{Adapting to fluctuating offline inference demand.}
Given their inherently relaxed SLOs, offline workloads offer an opportunity to leverage {CPU capacity during periods of low CPU utilization}.
EcoServe's reuse strategy is motivated by two key insights:
\begin{enumerate}
    \item {Offline inference demand is significant}: Figure~\ref{fig:offline} illustrates the ratio of online vs. offline demand in a real-world cloud deployment over a week (left) and a single day (right). On average, {offline workloads account for 21\% and 45\% of total infrastructure capacity} in Services A and B, respectively.
    \item {Offline workloads exhibit time-varying peaks}: At peak periods, offline demand reaches {27\% and 55\%}, respectively. By {scheduling offline inference to underutilized CPU nodes}, EcoServe {reduces peak GPU resource requirements}, thereby lowering total embodied carbon.
\end{enumerate}
Since offline workloads lack stringent latency constraints, EcoServe {batches large offline requests} on CPUs, thereby reducing the need for high-GPU provisioning.
For datacenters with high-carbon-intensity power grids, EcoServe can {dynamically reallocate workloads} back to GPUs when energy efficiency becomes a priority.

\begin{figure}[t]
    \centering
    \includegraphics[width=0.8\linewidth]{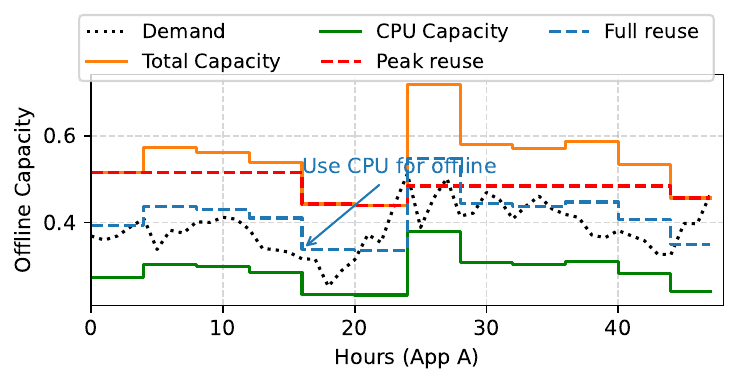}
    \vspace{-0.2in}
    \caption{Impact of CPU reuse strategies on infrastructure capacity. Red and blue curves represent peak-only and continuous reuse, respectively. EcoServe dynamically reallocates workloads to CPUs, reducing offline GPU capacity demands by up to 1.32$\times$.}
    \label{fig:offline-result}
    \vspace{-2em}
\end{figure}

\textbf{Load-aware reuse.}
To quantify the impact of CPU reuse, Figure~\ref{fig:offline-result} presents the required infrastructure capacity for serving Llama-3-8B inference workloads under varying demand conditions.  
We analyze two scenarios:
\begin{itemize}
    \item {Peak-aware reuse (red):} CPUs handle offline inference \textit{only} during peak demand periods.
    \item {Continuous reuse (blue):} CPUs process offline inference \textit{at all times}.
\end{itemize}
Assuming resource reallocation occurs every {4 hours, EcoServe’s load-aware CPU reuse reduces offline GPU provisioning needs by {1.32$\times$ at peak demand}, contributing to substantial {embodied carbon savings}.
This estimate is {conservative}, assuming fixed batch sizes for CPU and GPU inference. By further increasing CPU batch sizes, offline capacity reductions of up to 45\% are achievable.}

\subsubsection{\textbf{Right-sizing GPU Provisioning}} 
\label{sec:rightsize}
While CPU reuse amortized embodied carbon, efficient GPU provisioning is equally critical. EcoServe optimizes GPU allocation by considering three key factors: energy grid carbon intensity, workload characteristics, and hardware diversity. In regions with predominantly renewable energy, CPU reuse is prioritized; in fossil fuel-dependent regions, workloads may shift to more energy-efficient GPUs to minimize overall emissions.

LLM has distinct phases (prompting, decoding) and workload has various characteristics (architecture, sequence length)--this impose different compute and memory requirements. By dynamically matching these requirements to diverse GPU capabilities, EcoServe reduces both operational and embodied carbon while maintaining performance targets.

\textbf{Heterogeneous GPU partitioning.} 
Figure~\ref{fig:relative_carbon} compares the energy (bottom) and embodied carbon efficiency (top) of NVIDIA A100 vs. H100 for Gemma 27B across varying prompt lengths and batch sizes. The analysis separates prompt computation and decode (token generation)~\cite{patel2023splitwise}. A100 is preferable for smaller inputs and batches, while H100 becomes more efficient as they grow. However, for decode, A100 consistently outperforms H100 due to H100's low Model FLOPs Utilization (MFU), Memory Bandwidth Utilization (MBU), and high embodied carbon and energy cost.

We propose a workload-aware GPU provisioning strategy that optimizes resource allocation based on workload characteristics (not only on prompt/decode phases). Instead of provisioning only H100 for prompting (like what was done in Splitwise~\cite{patel2023splitwise}), we tailor GPU allocation per (workload slice, SLO) using offline profiling and performance modeling, distinguishing between workloads to match the available hardware heterogeneity.

\textbf{Exploiting tensor-model parallelism.}
The decision of how many levels of tensor parallelism (TP) is dependent on the target metrics. From latency perspective, sharding tensors across GPUs with high bandwidth communication links enables faster computation and more HBM bandwidth, it's generally favorable for large models to do higher level of TP (up to the tile and wave quantization effect~\cite{nvidia_matmul_guide}). However, the latency comes at a cost with more communication overheads. From carbon perspective, the TP level depends on the ratio between CPU embodied and GPU embodied carbon. Table~\ref{tab:tp} details the desiderata for TP level deployment. In practice, we tailor it to each hardware, model, batch size and SLO. %

\begin{figure}[!t]
\centering
    \begin{subfigure}[b]{\linewidth}
    \includegraphics[width=0.99\linewidth]{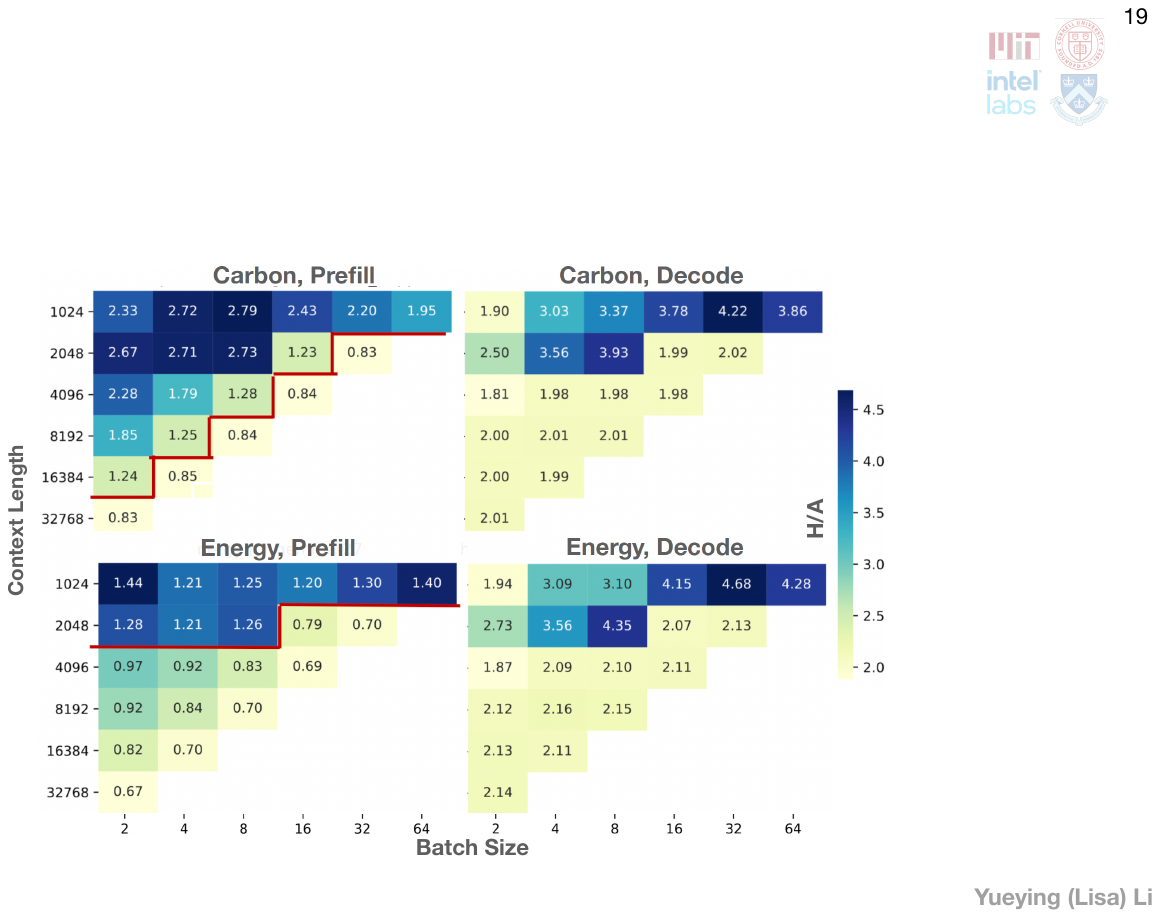}
        \label{fig:relative_carbon}
    \end{subfigure}
\vspace{-2em}    
\caption{Relative energy and carbon of prompt and decode for Gemma-27b model on NVIDIA H100 compared with A100 GPUs. The results are normalized (higher than 1 means  A100 is preferred). 
We find the optimal GPU varies based on LLM phase, context length, and batch-size, motivating the need for heterogeneous GPU \textit{rightsizing} for different workloads.}
\label{fig:relative_carbon}
\vspace{-1em}
\end{figure}

\begin{table}[h]
\centering \footnotesize
\begin{tabular}{ccc}\hline
\textbf{Metric} & \textbf{Relative Scale} & \textbf{Decision Criteria} \\ 
\hline
Power & $\frac{2n P_{GPU} + P_{CPU}}{n P_{GPU} + P_{CPU}}$ & Consider if $P_{CPU} \gg n P_{GPU}$ \\
Latency & $\approx 0.5 + C_{comm}$ & Favorable when model $>$ memory  \\
Cost/Model & $\approx 1$ & Near-constant when $C_{CPU} \ll n C_{GPU}$ \\
 Carbon & $\frac{CF^{emb}_{CPU} + n CF^{emb}_{GPU}}{CF^{emb}_{CPU}/2 + n CF^{emb}_{GPU}}$ & Better with higher $CF_{CPU}/CF_{GPU}$ \\
Energy & $\approx 0.5$ & Linear improvement with fixed CI \\
\hline 
\end{tabular} \caption{Power, latency, carbon, energy and cost when scaling up model and tensor-model parallelism.}\label{tab:tp}
\vspace{-2em}
\end{table}

\subsubsection{\textbf{Reduce: Host System Provision}} \label{sec:reduce}
This section examines strategies to minimize host memory subsystem resource underutilization to reduce embodied carbon waste. We focus on optimizing memory and storage resources, which comprise 36\% of embodied emissions in Azure A100 offerings\footnote{Standard\_ND96asr\_A100\_v4, the number is generally higher for higher capacity instances}. These optimizations must be balanced against the benefits of CPU reuse for offline inference, as explored in Section~\ref{sec:reuse}.

\textbf{Prefix-caching-aware DRAM reduction strategy. }
DRAM supports AI inference through:
\begin{itemize}
\item Memory offloading and intermediate data storage
\item Data processing and memory allocation
\item KV Cache swap space and prefix caching
\end{itemize}
For online inference with tight SLOs, offloading is not viable for large models. The minimum DRAM capacity is determined by:
\begin{equation}
\begin{gathered}
C^{\text{DRAM}} = C^{\text{Weight (layer)}} + C^{\text{KV/activation offload for online}} + C^{\text{KV for offline}} \\
\min C^{\text{DRAM}} = M_{k v}(n) = 4 n d h_{k v} l
\end{gathered}
\end{equation}
where $d$ is model dimension, $l$ is layer count, $n$ is P90 aggregated context length with zero reuse distance, and $h_{kv}$ is KV head dimension.
To mitigate performance impact from reduced DRAM, we use profiling to understand prefix reuse distances and determine optimal CPU cache space. We route longer-context requests to DRAM-abundant servers and leverage solutions like CacheGen~\cite{CacheGen} to offload caches across multiple instances or to SSD.

\textbf{KV-offload-aware SSD reduction strategy. }
While cloud GPU offerings typically include large SSDs, this isn't fundamental for inference serving. SSDs consume approximately 2.8~W per TB idle power (>10\% of server idle power). The minimum required SSD size approximates GPU memory capacity:
\begin{equation}
\begin{gathered}
C^{\text{SSD}} = 1.2 C^{\text{GPU}} + C^{\text{Model buffer}} + C^{\text{KV Offload for offline}} \\
\min C^{\text{SSD}} = 1.2 C^{\text{GPU}}
\end{gathered}
\end{equation}
When SSD capacity is used to download and maintain models, we imagine remote storage with GPUDirect can reduce over provisioned storage.

\subsubsection{\textbf{Recycle: Asymmetric Lifetime for Host and GPUs.}}\label{sec:recycle}
Due to the nature of fast energy efficiency improvements of GPU and slow energy improvement on hosts, the optimal lifetime $T$ is also different, necessitating an asymmetric recycling strategy.

\textbf{Host lifetime extension.}
Extending the lifespan of DRAM, SSDs, and CPUs helps reduce both embodied and operational carbon. This recycling opportunity arises because memory and storage bandwidth have not scaled proportionally with CPU/GPU performance in recent technology nodes. Our measurements across three server generations show that DRAM and SSD bandwidth are rarely the primary bottlenecks in modern inference serving systems, allowing older components to be reused without compromising performance.

\begin{figure}[t]
\centering
     \includegraphics[width=0.99\linewidth]{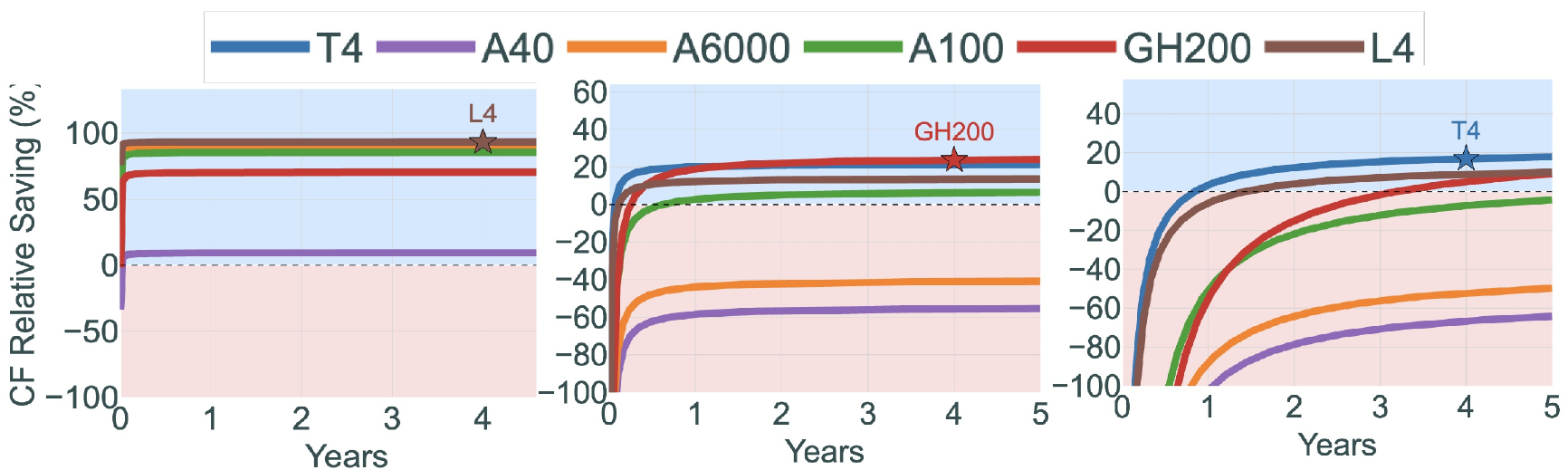} \vspace{-1em}
    \caption{Relative carbon savings with various hardware compared with V100. (Left) prompt heavy workload with Carbon Intensity (CI)  400 gCO2e/KWh, (Middle) prompt heavy workload or  (Right) decode heavy with CI = 50 gCO2e/KWh, The optimal hardware is different. }\label{fig:cf-relative}
    \vspace{-1.05em}
\end{figure}

\textbf{Carbon-aware GPU upgrade.} The GPU upgrade strategy depends on 1) inter-generational GPU energy efficiency improvement, 2) the workload and model characteristics, and 3) the upfront embodied carbon cost. Upgrading from V100 to GH200 incurs an upfront embodied carbon cost but significantly improves token processing efficiency per unit of operational carbon. Over time, this reduces the relative carbon footprint, especially in datacenters with lower renewable energy availability. Figure~\ref{fig:cf-relative} shows the optimal GPU usage duration across regions with varying carbon intensity.

\begin{figure}[h]
    \centering
    \includegraphics[width=0.8\linewidth]{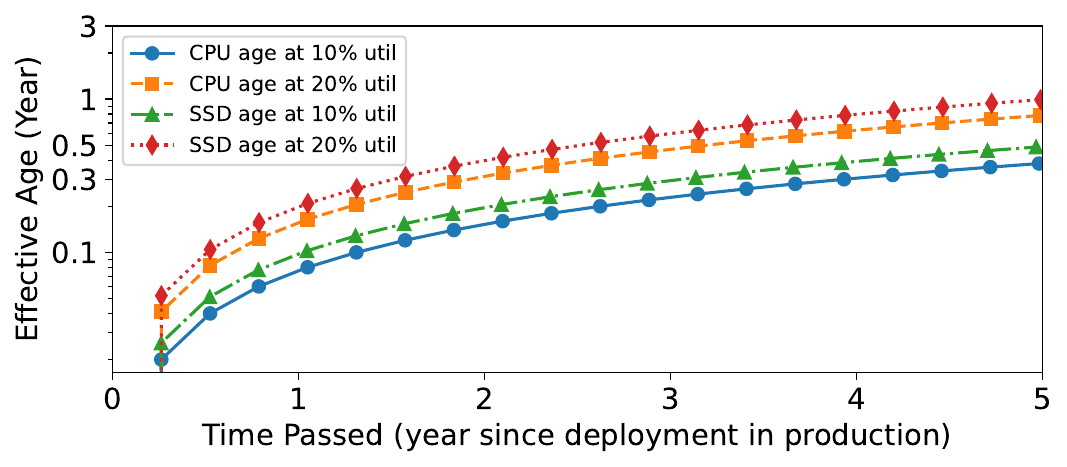}
     \vspace{-1.05em}\caption{Effective age with deployment time.}
    \label{fig:reliability} \vspace{-1.05em}
\end{figure}

\textbf{Reliability implications. } Production data indicates that AI inference workloads typically utilize these host components at low levels. 
 In this section, we study how utilization and lifetime impact reliability. 
1) CPU: To understand the CPU aging over 5 years in a cloud setup, we use a 7nm composite processor model from a major fabrication company. The model faithfully represents CPU transistor logic and is used for design and compliance with reliability goals; details omitted due to confidentiality. In the usual spread of observed voltage levels, we found that even at 20\% utilization over 5 years, CPU aging is only 0.8 years, indicating significant potential for extended use (Figure~\ref{fig:reliability}).

2) DRAM: A study~\cite{Siddiqua2017} from a heavily utilized Cielo supercomputer showed that even at the end of lifetime of 5 years, the DRAM error rate did not increase, indicating a higher lifetime. Other recent studies have shown that DRAM retention rate based errors only meaningfully increase after 10 years of intense usage~\cite{liu2022new}. Given the low utilization of DRAM in the cloud~\cite{kanev2015profiling}, reusing old DRAM and SSDs in CPU-only servers has been proposed by prior work as well~\cite{greensku}, but EcoServe is the first to consider how to evaluate optimal and asymmetric lifetime extension. 
 
3) SSD: Age of an SSD primarily depends on the number of program/erase cycles the SSD has been through~\cite{failure-ssd,FlashFailure,schroeder2016flash,klein2021backblaze}. Therefore, an SSD’s age is modeled as proportional to the number of writes in its lifetime. Even if we assume that the SSD is written all the time that the CPU is active, it will only age by 1 year over a period of 5 years for a 20\% utilization (shown in Figure~\ref{fig:reliability}).

\subsection{Putting it all together} \label{sec:ILP}
Deploying EcoServe at scale requires balancing multiple competing objectives across carbon, energy, performance, cost, and reliability. The interactions between our optimization strategies present complex trade-offs. For instance, aggressive CPU \textit{Reuse} may conflict with \textit{Reduce} goals, requiring careful capacity planning for CPU cores and memory subsystems.

To address these challenges, we propose a hierarchical approach:

\begin{enumerate}
    \item At the capacity planning phase, we provision a spectrum of server configurations optimized for different carbon-performance trade-offs based on offline profiling. Some servers emphasize \textit{Reuse} with expanded memory subsystems, while others focus on \textit{Reduce} with lean configurations.
    \item At runtime, a carbon-aware load balancer leverages these diverse configurations to optimize workload placement, request scheduling and resource allocation. The load balancer considers:
    \begin{itemize}
        \item Server configuration profiles
        \item Real-time workload characteristics and system load
        \item Dynamic KV cache transfer costs
    \end{itemize}
\end{enumerate}

\subsubsection{Offline profiling and online adaptation}
As shown in Figure~\ref{fig:ecoserve}, EcoServe optimizes LLM deployment by ingesting models, workload traces, SLOs, and hardware specs. It bootstraps initial deployment using CPU/GPU profiling data, then continuously adapts to changing workload patterns. The framework maintains separate resource pools for online, mixed and offline inference, with dynamic CPU offloading for decode phases. Pool sizes automatically adjust via periodically triggered ILP based on workload demands and carbon intensity.%

\subsubsection{ILP for co-design scheduling and allocation}

\textbf{Workload Slicing and Disaggregation}: We represent the workload in different slices within each phase {prompt} and {decode}. Let \( H(i,o) \) represent the request rate for inputs of length \( i \) and outputs of length \( o \). We put workload into histogram bucket \( b \), and further divided into \( S \) slices for fine-grained allocation, where slice factor \( f \) determines the granularity. For a bucket with request rate \( \lambda_b \), each slice \( s \) has rate \( \lambda_s = \lambda_b / f \). For each slice \( s \in S \), we define $\lambda_s$ as the request rate, $\text{Lat}(s,g,l, \phi_s,m_s)$ as expected latency for the prompt or decode phase on GPU \( g \) with allocated resources at load \( l \), required CPU cores \( \phi_s \), and memory \( m_s \). The load imposed by slice \( s \) on GPU type \( g \) is calculated separately for each phase:
\[\footnotesize
\text{Load}_{p/d}(s,g) = \frac{\lambda_s}{\text{MaxTput}_{p/d}(g,\text{size}(s),\text{SLO})}
\]
where \( \text{size}(s) \) represents the input/output length pair of slice \( s \).

\textbf{Performance and Carbon Model}: For each GPU type \( g \in G \), we define \( c_g,  c_\phi, c_m \) as hourly GPU, CPU, memory cost, \( \gamma_g(t) \) as GPU carbon intensity at time \( t \), which is both linear with operational power of the workload times the carbon intensity, and with unit-time embodied carbon of the hardware.
The carbon impact of a workload is the sum of both phases $\text{Carbon}(s,g,l, \phi_s,m_s) = \sum_{k \in \{p,d\}} \gamma_g(t) \cdot \text{Lat}_{k}(s,g,l,\phi_s,m_s)$.

\textbf{ILP Formulation}: Our goal is to minimize the weighted carbon cost and latency-dependent hardware cost under service level objectives (SLO) based on both TTFT (time to first token) and TPOT (time per output token). Decision variables include GPU assignment matrix \( A \in \{0,1\}^{N \times M} \), GPU count vector \( B \in \mathbb{Z}_{\geq 0}^M \); CPU cores and memory allocated to slice \( s \) as \( \Phi_s, M_s  \).
\label{sec:optimization}
{\footnotesize
\[
\min_{A,B,\Phi,M} (1-\alpha) \Big[\sum_{j=1}^M B_j c_j + \sum_{s=1}^{N} (\Phi_s c_\phi + M_s c_m) \Big] + \alpha \sum_{s=1}^N \sum_{j=1}^M A_{s,j} \text{Carbon}(s,j,l, \Phi_s,M_s)
\]
\vspace{-2.em}
\begin{align*} 
\textbf{s.t.}\quad & \forall s, \quad \sum_{j=1}^M A_{sj} = 1 \\
& \forall j, \quad \sum_{s=1}^N A_{s,j} \cdot (\text{Load}_{p}(s,j) + \text{Load}_{d}(s,j)) \leq B_j \\
& \sum_{s=1}^N \Phi_s \leq \Phi, \quad \sum_{s=1}^N M_s \leq M \, \quad \forall s, \quad \Phi_s \geq \phi_s, \quad M_s \geq m_s \\
& \forall s,j \text{ where } A_{s,j} = 1, \quad \text{Lat}_{p/d}(s,j,l, \Phi_s,M_s) \leq \text{SLO}_{p/d}
\end{align*}
}\vspace{-2em}

where $\alpha \in (0,1)$ is a weighting variable between cost and carbon objectives. Unless otherwise stated, we choose $\alpha = 1$. A smaller value such as $\alpha = 0$ will simply optimize for cost.

\section{Implementation and Experimental Setup}

\textbf{Hardware Systems.}
The experimental evaluation uses a heterogeneous set of CPU and GPU hardware.
In terms of GPUs we use a combination of PCIe-based NVIDIA H100, A100, A6000, L4, and A40 GPUs, representing a diverse range of compute, memory, power, and energy-efficiency tradeoffs.
In terms of CPUs, we use a dual-socket server class Intel Sapphire Rapids CPU. The CPU frequency governor is set at performance mode with turbo boost disabled. %

\textbf{Models, Workloads and Load Generator.} 
We evaluate our system using a wide range of workloads to capture diverse scenarios. 
For models, we assess different architectures with varying sizes. For request generator, we use scaled Azure function traces (AZF) and production offline workload traces~\cite{azure_public_dataset} (Figure~\ref{fig:offline}). For workload size, we sample from public datasets listed below: 
\[\footnotesize
\begin{array}{cccc}
    \toprule
    \textbf{Model} & \textbf{TTFT} & \textbf{TPOT} & \textbf{Dataset} \\
    \midrule
    \text{Gemma-2-2B} & 0.25 \, \mathrm{s} & 0.1 \, \mathrm{s} & \text{ShareGPT~\cite{sharegpt}} \\
    \text{MetaLlama-3-8B} & 0.5 \, \mathrm{s} & 0.1 \, \mathrm{s} & \text{ShareGPT~\cite{sharegpt}} \\
    \text{Llama-13B} & 1.5 \, \mathrm{s} & 0.15 \, \mathrm{s} & \text{AFT~\cite{azure_public_dataset}} \\
    \text{Llama-70B} & 15 \, \mathrm{s} & 0.24 \, \mathrm{s} & \text{AFT~\cite{azure_public_dataset}} \\
    \text{Mixtral-8x7B} & 2.5 \, \mathrm{s} & 0.15 \, \mathrm{s} & \text{ShareGPT~\cite{sharegpt}} \\
    \text{(Online) Gemma-2-27B} & 10 \, \mathrm{s} & 0.2 \, \mathrm{s} & \text{AFT~\cite{azure_public_dataset}} \\
    \text{(Offline) Gemma-2-27B} & 24\mathrm{hr} &  \, \mathrm{N/A} & \text{LongBench~\cite{longbench} } \\
    \text{Bloom-176B} & 20 \, \mathrm{s} & 0.27 \, \mathrm{s} & \text{AFT~\cite{azure_public_dataset}} \\
    
    \bottomrule
    \end{array}
\]

\textbf{Software.} We use vLLM v0.6.2 and Llama.cpp (d5ed2b9) to deploy LLM instances across provisioned GPU and CPUs. We also use Splitwise simulator~\cite{patel2023splitwise} and integrate our carbon models to evaluate the comparison and impact of more fine-grained provision strategy (Section~\ref{sec:sensitivity}) of different scales. We implement the profiling-based performance, energy and carbon models using Python based on those frameworks under various hardware, workload traces and request generators. We implement the ILP formulation and resource allocation algorithms based on CVXpy~\cite{cvxpy}. We sample from Watttime~\cite{watttime} and GreenSKU~\cite{greensku} to get the real-time or average carbon intensity of various regions. The aggregated analysis in Figure~\ref{fig:result-summary} Left reflects  the average carbon intensity at mid level (261 gCO2/kWh). %

\section{Evaluating EcoServe}

We evaluate EcoServe across three dimensions: carbon efficiency, performance, and energy. Our evaluation focuses on the following key research questions:

\begin{itemize}[leftmargin=*]
    \item How much additional operational and embodied carbon can EcoServe save in comparison to performance, energy, and cost-optimized baselines?
    \item How does EcoServe perform to different carbon intensity and request rates?
    \item What's the overhead and scalability of the core resource allocation components in EcoServe?

    \item What are the carbon footprint implications of combining multiple categories of EcoServe's optimization strategies? What are the tradeoffs with performance? 
    \item What is the effectiveness of individual design strategy in 4Rs?
\end{itemize}
\subsection{End-to-end Evaluation}

\begin{figure*}
  \centering
    \centering
    \begin{subfigure}{0.6\textwidth}
\centering
    \includegraphics[scale=0.5,viewport=150 20 550 160]{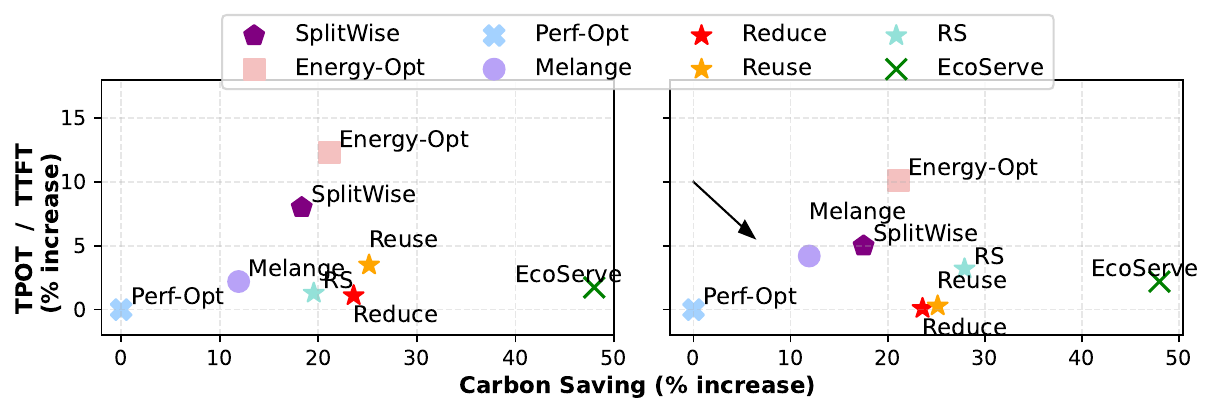}\vspace{-0.03in}
\vspace{-0.03in}
    \end{subfigure}
    \begin{subfigure}{0.38\textwidth}
      \centering
      \hspace{-0.9in} 
       \includegraphics[width=1.1\linewidth]{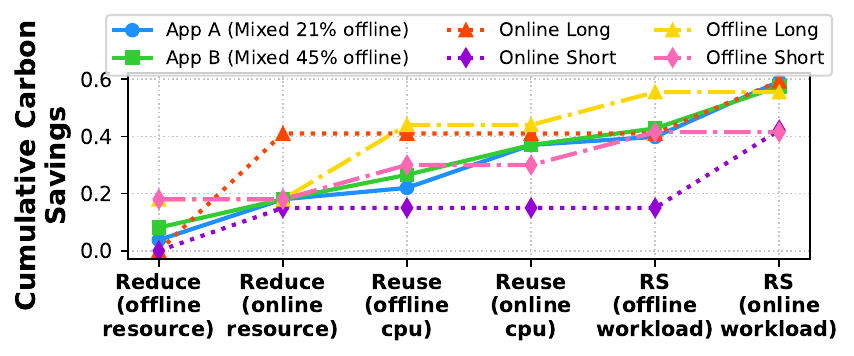}\vspace{-0.08in} \label{fig:decompose-4R}
    \end{subfigure}
    \caption{Carbon vs. performance trade-off. Lower-right is better. (Left, Center) We compare EcoServe to performance, energy, and cost (Melange) optimized baselines for prompt latency (TTFT, center) and generation latency (TPOT, left). We show the individual improvements (i.e., reuse, reduce, rightsize, recycle) as well as the aggregated benefit.
    EcoServe achieves up to 47\% carbon reduction at comparable performance.
    (Right) Under different workload mixes, we show the cumulative benefits of successively applying EcoServe's optimizations to online and offline capacity. All experiments assume iso-throughput.}
    \vspace{-1em}
    \label{fig:result-summary}
\end{figure*}

We begin by summarizing the overall carbon and performance implications of EcoServe.
Figure~\ref{fig:result-summary} (left) demonstrates the average relative performance degradation and carbon improvements normalized to performance-optimized (\textit{perf-opt}) system configuration and resource allocation baselines. 

The baselines are configured as follows: (1)
a performance optimized (\textit{perf-opt}) configuration implements a single hardware that minimizes TTFT and TPOT, 
(2) a cost-optimized baseline (\textit{Melange}) which provisions hardware to minimize cloud cost, 
(3) an energy-optimal configuration (\textit{energy-opt}) that finds software provision (GPU resource allocation) decisions to minimize energy (with no capacity planning changes on CPUs), (4) \textit{SplitWise} -  provisioning strategy for H100 and A100, using the simulator to pick the best carbon configuration on Pareto frontier under SLO constraints.  
We drive the workload with a Poisson request generator with various arrival rates and scaled Azure Function Traces (2023) to emulate the bursty behavior of online samples (5\% - maximum system load).

Figure~\ref{fig:result-summary} (left,middle) shows that EcoServe variants can achieve significant carbon savings compared with performance optimal configurations for both online and offline workloads. We further decompose the carbon savings from various strategies:

\begin{itemize}[leftmargin=*]
    \item Reduce can yield around 12.4-28.6\% carbon savings. For leaner GPU offerings like T4, the savings are less than higher-end GPUs since the host is designed to scale with the GPU memory capacity.  
    \item Right sizing (RS) enables around 15.2-30.3\% better carbon, leveraging the SLO slacks for various workloads on TPOT and TTFT. The benefit is particularly significant in lower carbon intensity regions with higher load variability and higher online workload component, as EcoServe can dynamically adjust its resource allocation to match demand to different workload slices. 

    \item Optimized CPU reuse strategies can save up to 25.4\% of carbon for clusters with different percentages of offline workload. 
    \item Recycle strategy is orthogonal to all the other Rs. It can be independently deployed, yielding around 16.8\% savings compared with a homogeneous update baseline (Section~\ref{sec:recycle-eval}).
    \item When it comes to combining three strategies together, RS + Reuse are software solutions that produce more synergy, whereas Reduce and Recycle are orthogonal hardware design optimizations. On average EcoServe yields 47\% carbon savings. 
\end{itemize}
The results motivate co-designing the 4Rs for different workloads. 
\subsubsection{Performance Analysis}
\begin{figure}
    \centering
    \includegraphics[width=0.65\linewidth]{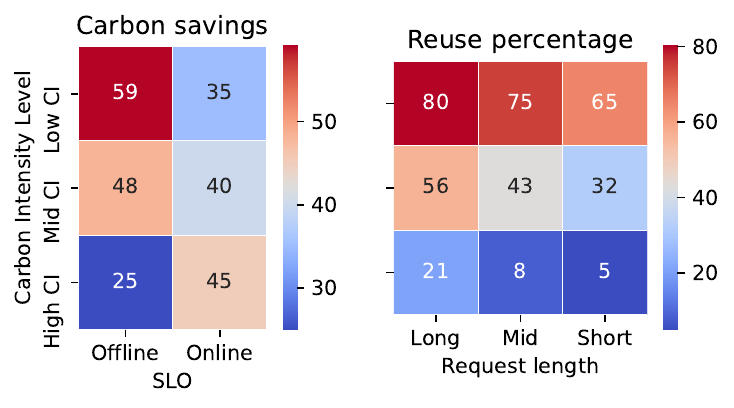}\vspace{-0.23in}
    \caption{Carbon savings and reuse configurations sampled by EcoServe under various workload lengths, SLOs and carbon intensities for a mid-sized model (Llama-70B). }
    \label{fig:ablation}
\end{figure}

From Figure~\ref{fig:result-summary}, we demonstrate that the performance of EcoServe variants is close to Perf-opt within 3 percent, except for the reuse case. The reason for a worse TPOT (around 5\%) is because of the offline workloads, where the reuse strategy prioritize the decoding of tokens with longer context length to be offloaded to CPU. The KV cache transfer This performance drop can be recovered with better hardware interconnects and other system techniques like batching transfer for smaller models to improve MBU (memory bandwidth utilization).

EcoServe (RS) provides better improvements on TPOT metric than TTFT metric in the presence of SLO slack.
This is because of the longer prompt and higher load, when we right-size the allocation periodically, the unavailability will cause the prompt scheduler to queue some requests' prompts temporarily, which results in the queuing delay. Furthermore, by design, \textit{Reuse} mostly addresses the \textit{decoding servers'} capacity saving, hence the performance on TTFT on the \textit{prompting servers} is untapped. EcoServe (Reduce) incurs a little overhead on TPOT (<5\%), but the \textit{energy-opt} allocator is worse in performance through aggressively choosing energy-opt hardware (like L4), ignoring the embodied carbon per performance. Furthermore, \textit{Splitwise} suffers in carbon and performance because of the limited hardware options, and the gap between provision and scheduling decisions (Section~\ref{sec:rightsize}). Further analysis on a specific workload is shown in Figure~\ref{fig:split-compare-2}.

\subsubsection{Breakdown Analysis.}

The benefits of carbon footprint reduction in EcoServe can be uncovered layer by layer through various strategies (Figure~\ref{fig:result-summary} Right).

For online workloads, \textit{Reduce (online resource)} allows us to save approximately 17-40\% of the carbon footprint by reducing DRAM and SSD capacity. When moving to \textit{Reuse}, online workloads seldom use CPU resources, maintaining consistent savings. The \textit{RS (online)} strategy further increases savings by employing workload-aware load balancers and resource allocators, limiting the overprovisioning of high-end hardware.

For offline workloads, savings from the \textit{Reduce} strategy are less significant. The \textit{Reduce (offline resource)} method is conservative, requiring DRAM to match the model size for offline inference on CPU. However, substantial savings come from reusing offline CPU capacity for decoding tasks. Rightsizing also yields significant benefits due to relaxed SLO and the Integer Linear Programming (ILP) formulation to optimize GPU sets through disaggregation. Longer offline workloads yield more benefits. 

For mixed workload production traces, workload (A), which has less offline demand, exhibits less benefit from aggressive \textit{Reduce (online)}; however, the overall carbon savings stay relatively similar for the two mixed workloads based on our analysis, demonstrating the generality for our approach across different types of workloads.

\subsection{Sensitivity and Scalability Study} \label{sec:sensitivity}

\subsubsection{Varying Carbon Intensity.}
To show the robustness of EcoServe's carbon savings under various geographical regions, models,, and workload patterns with different carbon intensity levels, we conducted a controlled experiment based on Splitwise code traces (for 2 minutes) on the simulation infrastructure for a medium/large model (Llama-70b, Bloom-176b) with different request arrival rates. 

\begin{figure}[t]
    \centering
    \includegraphics[width=0.9\linewidth]{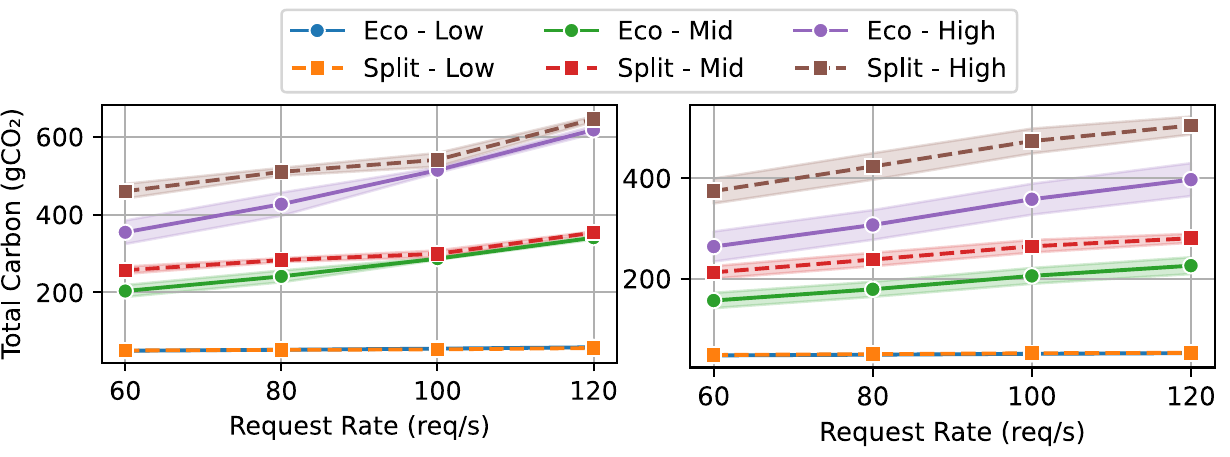}\vspace{-0.2in}
    \caption{Comparison of Bloom-176B (Left) and Llama-70B (Right) with Splitwise on various carbon intensities (Low, Mid, High) and load on iso-power deployment. EcoServe consistently outperforms Splitwise, and the gap is larger under a lower request rate and higher CI. }
    \label{fig:split-compare-2}
\end{figure}

We selected three power
grids: North Central Sweden (Low), California (Mid), and
Midcontinent (High). These regions represent low, medium,
and high carbon intensity levels, with carbon intensities of
17, 261, and 501 gCO2/kWh, respectively. The total carbon is shown in Figure~\ref{fig:split-compare-2}. Notably, EcoServe selects the software provision strategy that is different from SplitWise because the runtime could allocate shorter prompts to the A100. This leads to carbon savings on that trace for 26.5\% on average. 

Under a 40 H100 power-equivalent setup with
long input sequences, Splitwise achieved Pareto-optimal energy-throughput
by allocating 35 Prompt and 8 Token machines. However, its JSQ scheduling
missed workload-aware co-design opportunities. EcoServe matched throughput
with 30 H100s and 17 A100s by adaptively placing longer requests to
A100s, reducing both embodied and operational carbon by up to 38\% and 36\% for large and medium models under lower load. 

From Figure~\ref{fig:ablation}, we observe that for longer requests, lower carbon intensity level, reuse is most often used. The carbon savings for offline workload is most significant. However, when carbon intensity is higher, we save more with \textit{right-sizing} and \textit{reduce} strategy for online workloads.

\subsubsection{Scalability of the control plane}
\begin{table}[b]\footnotesize
    \centering
    \begin{tabular}{ccccc}\hline
        cluster size & online (low) & offline (low)  & online (high) & offline (high) \\ \hline
        10  & 0.281 & 0.343 & 0.428 & 0.523 \\
        20  & 0.377 & 0.521  & 0.574 & 0.789 \\
        40  & 0.398 & 0.672  & 0.572 & 1.030\\
        80  & 0.508 & 0.883  & 0.747 & 1.272 \\
        160 & 0.585 & 0.919  & 0.901 & 1.315\\\hline
    \end{tabular}
    \caption{Overhead of control plane over different cluster sizes, workloads and system loads (Unit: second).}
    \label{tab:scalability} \hspace{-1in}
\end{table}
To show the scalability of EcoServe's control plane (ILP solver) with various cluster sizes and request arrival rates (low/high), we measured the time taken for the ILP solver and scheduler to make the right-sizing decision. 
Table~\ref{tab:scalability} shows the control plane overhead across different cluster sizes and load conditions. The overhead grows sub-linearly with cluster size, increasing by only 2.1-2.7$\times$ when scaling from 10 to 160 nodes. This favorable scaling comes from the ILP formulation's ability to exploit our carbon and latency performance model - similar workloads on similar hardwares are clustered together. Under high load, offline planning exhibits the highest overhead (1.315s at 160 nodes) due to exploring more complex hardware combinations. However, even in this worst case, the sub-2-second latency remains practical for production deployments where planning decisions typically operate on minute- or hour-level intervals. Online decisions show better scaling (0.585s at 160 nodes) by pruning out irrelevant spaces, focusing on immediate resource availability rather than long-term optimization.

\subsection{Reuse Strategy Evaluation}
\begin{figure}[h]
    \centering
    \includegraphics[width=0.9\linewidth]{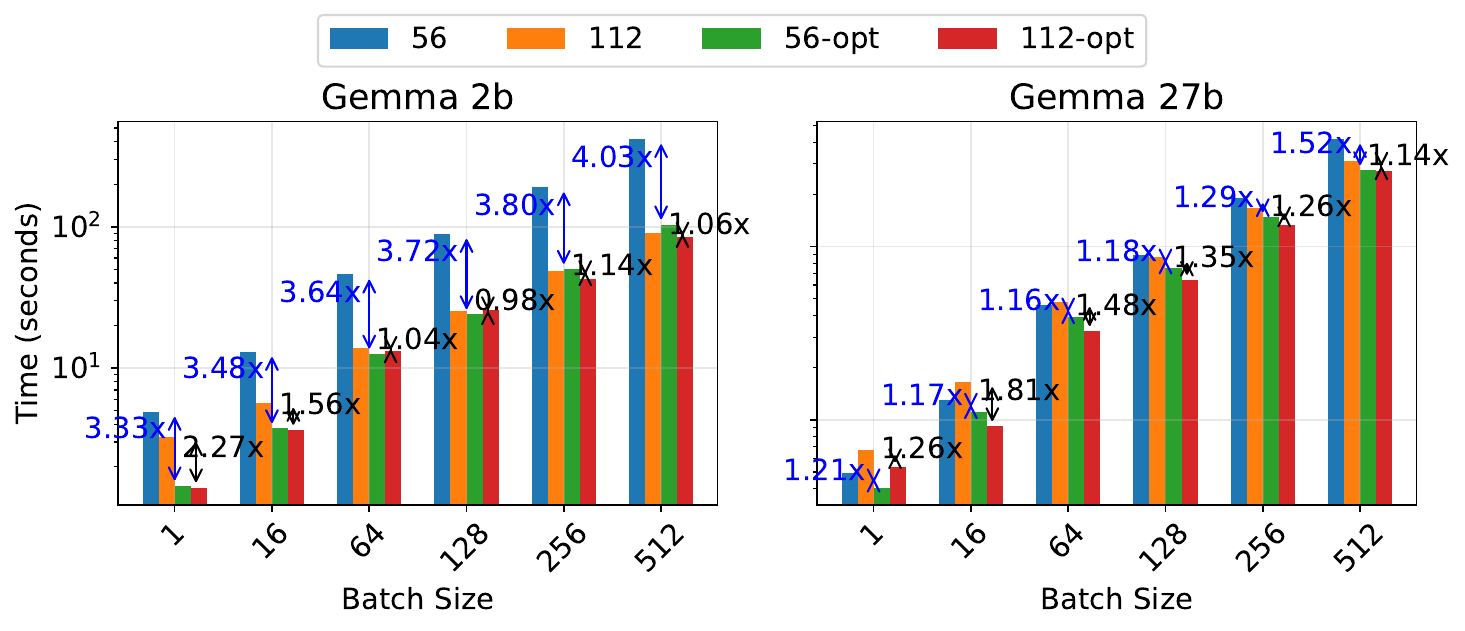}
    \vspace{-1.5em}
    \caption{Beginning with llama.cpp as a baseline, we optimize CPU decoding. EcoServe's CPU optimization improves performance by up to 4.03$\times$ and on average by 1.34$\times$ (across batch sizes and parallelism dimensions) on an Intel SPR.}
    \label{fig:speedup-reuse}
    \vspace{-1em}
\end{figure}

We evaluate the two design aspects of the reuse strategy: first, optimized CPU threading and tiling implementation for GEMM and long-context decoding. Second, workload- and load-aware CPU reuse strategy.
Figure ~\ref{fig:speedup-reuse} shows the speedup with various workload sizes with EcoServe's optimized \textit{reuse} strategy, compared with state-of-the-art CPU inference engine (llama.cpp) on Gemma 2/27B. The evaluation varies the core counts (56 vs 112) to show the generality of the approach. By choosing the right tiling and parallelism strategy guided by performance modeling, we improve decoding latency, which directly translates to higher throughput and carbon savings of 1.34$\times$ on average.

\begin{figure}
    \centering
    \includegraphics[width=0.9\linewidth]{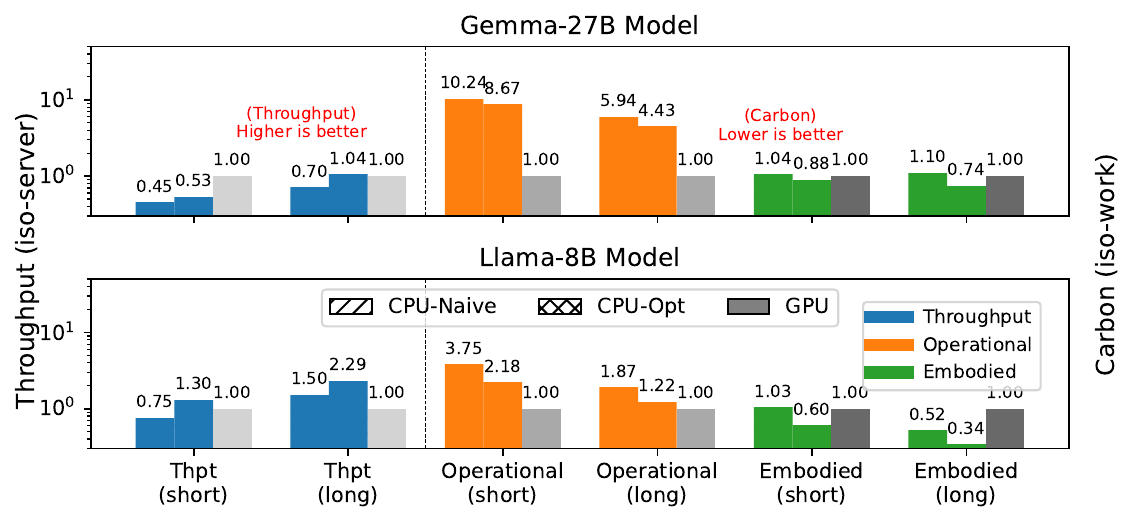}
    \vspace{-1.8em}
    \caption{Decoding throughput (Left), operational (Middle), and embodied carbon (Right), for various models and workload sizes, comparison with a naive llama.cpp vs. optimized CPU \textit{reuse} strategy, normalized to A100 GPU performance under maximum throughput. Carbon reduction is 12, 26\% for Gemma-27B, and 40\%, 66\% for Llama-8B, respectively, under short and longer workloads. }
    \label{fig:reuse-result}
    \vspace{-1.5em}
\end{figure}

Figure~\ref{fig:reuse-result} analyzes the detailed embodied and operational carbon breakdowns compared with serving on GPUs for EcoServe's \textit{reuse}, showing the benefit of workload- and load-aware optimization. Given the CPU's lack of energy proportionality, the added operational power is relatively minor but still results in operational carbon cost with longer execution time. The free lunch from a 56-core SPR machine CPU attached to A100 allows us to achieve 0.53-2.29$\times$ the throughput of the GPU (Blue bars). Although CPU generally has worse energy efficiency, especially for larger models when both CPU and GPU are compute bound (Orange bars), EcoServe's CPU decode can achieve up to 66\% relative embodied carbon savings compared with a naive llama.cpp (CPU-Naive) strategy as 48\%. On average, EcoServe can achieve 3.51$\times$  embodied carbon savings compared with a non-optimized baseline across the above workloads. More importantly, without EcoServe's \textit{reuse} strategy (CPU-Opt), the embodied carbon for llama.cpp (CPU-naive) is worse than non-reuse (GPU) for Gemma-27B model and Llama-8B model short context cases. 
This is due to the fact that under a setting with a longer context and smaller model dimensions, the gap between CPU and GPU throughput is larger due to the batching effect, and the carbon saving is more significant in an iso-throughput setting.

 The total carbon benefit arises from the net effect of embodied and operational carbon through \textit{Reuse}. EcoServe enhances cost and embodied carbon by utilizing unused CPU cores, reducing the need for over-provisioned GPUs. Additionally, it offers advantages in long-context decoding through customized CPU kernel optimization and allows flexible carbon reduction via runtime scheduling based on workload sizes.  However, it may not work well for shorter-context larger-model decoding in terms of operational carbon. The strategy can be tailored according to the region's carbon intensity. In areas with high carbon intensity (CI), operational carbon is a primary concern, so we should minimize CPU offloading. Conversely, in regions with low CI, embodied carbon plays a larger role, allowing us to increase capacity through CPU offloading (Figure~\ref{fig:ablation). }

\begin{figure}[h]
    \centering
    \includegraphics[width=0.89\linewidth]{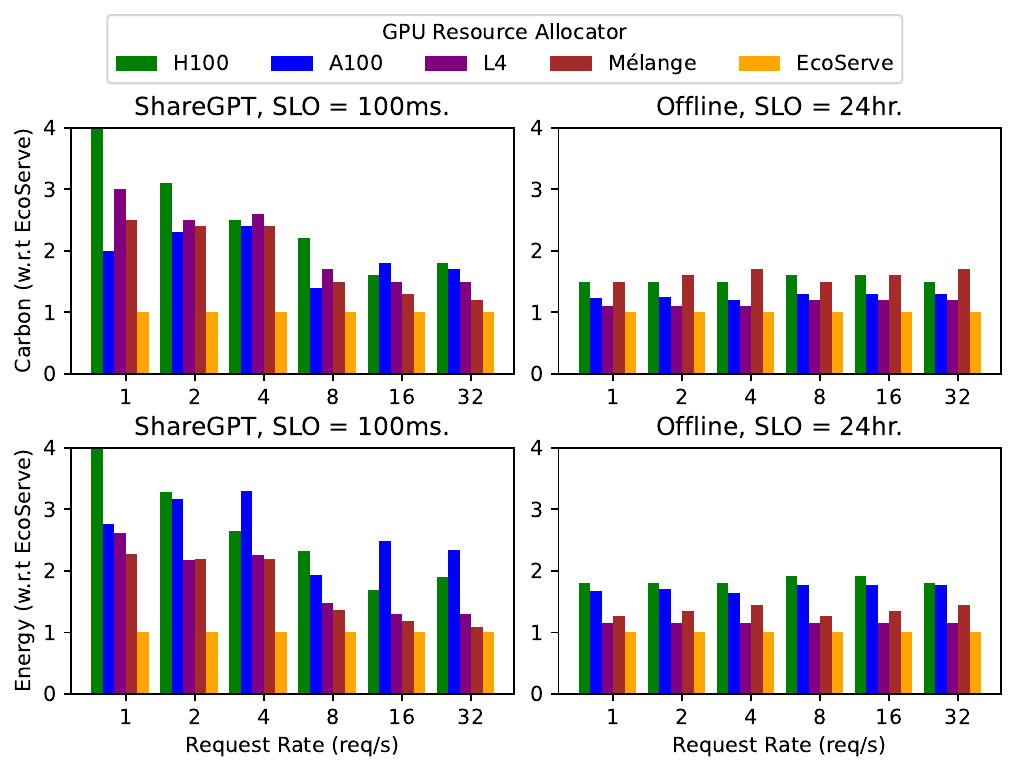}
    \vspace{-1.5em}
    \caption{Relative carbon and energy improvements of rightsizing GPUs on Gemma-27B on homogeneous GPUs (i.e., H100, A100, L4), and cost-optimized provisioning (Melange). EcoServe's \textit{Right-sizing} strategy exploits LLM phases to reduce overall energy and carbon under TPOT = 100 ms (online) and TTFT = 24hr offline constraints.}
    \label{fig:result-RS}
    \vspace{-1em}
\end{figure}
\subsection{Right-sizing Strategy Evaluation}
For right-sizing, we evaluate our system on Gemma-27B with various load and workload against the following two types of baselines: 1) Melange: Cost-optimal GPU resource allocation. 2) Single hardware: pick the single hardware and its replica numbers to satisfy workload demands that disregard SLO slacks and the energy/carbon implication for hardware choices. 

Figure~\ref{fig:result-RS} shows the final outcomes for \textit{right-sizing} strategy given different workload types and loads of the system. Each request generator runs for 5 minutes length traces, and we discard the first and last 5\% of the statistics for warming up effects. The length of the workloads is randomly sampled from the open-sourced traces with both long and short prompts.  %

For the online setting (left), under a lower request rate, EcoServe consistently outperforms Melange by over 2.56$\times$  for carbon and 2.25$\times$  for energy. 
The improvement comes from separating resource allocation decisions for LLM phases; finer-grained allocation enables better utilization and energy. As an example, for request rate of 1, EcoServe will choose the most carbon-efficient hardware based on the workload size; e.g. for medium-sized workload it chooses L4 and A100 for decoding and prompting, respectively; and A100 and H100 for long prompt. 
However, Melange sticks to the best perf-per-cost hardware for both phases for each workload slice, and the quantization effect for available hardware capacity introduces more over-provisioned resources for decode phase, which negatively impacts embodied and operational carbon.

EcoServe also performs better than other single hardware choices both on carbon and energy. The closest baseline is L4 due to its higher energy and carbon efficiency. This is because of the flexibility in allocating a combination of resource based on the optimal energy or carbon profile for different workload slices under various SLO. 

Furthermore, for offline setting (offline), EcoServe consistently wins in both energy and carbon up to 1.6$\times$ with the next best baseline, as it considers both dimensions in the optimization objective.
Under the carbon intensity of US-Central / South, operational carbon is around 0.64$\times$ of the embodied, making EcoServe a desirable solution for balancing both energy and embodied carbon. 

\subsection{Recycle Strategy Evaluation} \label{sec:recycle-eval}

\begin{figure}
    \centering
    \includegraphics[width=0.7\linewidth]{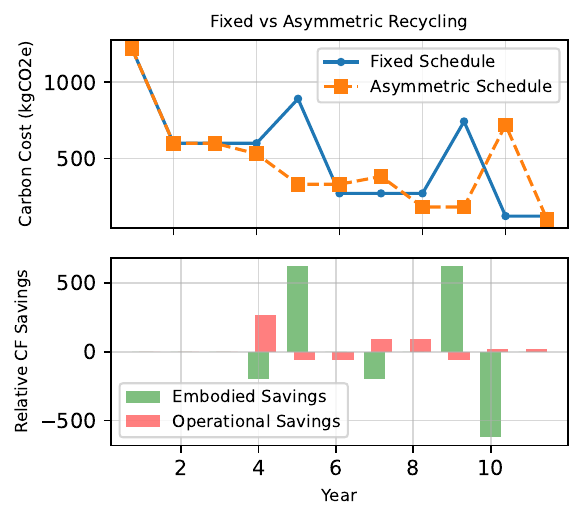}
    \vspace{-1.2em}
    \caption{Trends in annual embodied and operational carbon based on EcoServe's asymmetric recycling of CPUs (9 years) and GPUs (3 years). Baseline is a fixed schedule for both CPU and GPU (4 years). }
    \label{fig:speedup-recycle}
    \vspace{-2em}
\end{figure}

Figure~\ref{fig:speedup-recycle} shows absolute carbon cost (top) and relative savings (bottom) for fixed (baseline) vs. asymmetric (EcoServe) upgrades. The baseline assumes 800 kgCO$_2$e for the host, 120 kgCO$_2$e for the GPU, and 600 kgCO$_2$e in yearly operational emissions. GPU energy efficiency is estimated to double every 3.5 years on average~\cite{sun2019summarizing}. The baseline upgrades both host and GPU every 4 years, while EcoServe upgrades hosts every 9 years and GPUs every 3 years to better match efficiency gains.

The bottom plot demonstrates the relative carbon savings enabled by the asymmetric strategy. By upgrading host systems less frequently, significant embodied carbon savings are achieved due to their higher upfront embodied carbon costs. Although this can cause some negative operational carbon savings, host systems have less impact on overall operational emissions to AI inference compared to GPUs.
At the same time, more frequent GPU upgrades, which incur relatively lower embodied carbon costs, allow for earlier adoption of energy-efficient hardware, leading to operational carbon savings.
This trade-off enables EcoServe to balance embodied and operational carbon reductions.
Over the 10-year period, EcoServe asymmetric strategy is able to achieve approximately 16\% cumulative carbon saving compared to the baseline fixed strategy. 

\section{Related Work}

\textbf{Accelerating CPU Inference:} LLM inference on CPUs, as demonstrated by frameworks like Llama.cpp and Gemma.cpp, has garnered significant attention for enabling cost-effective deployment~\cite{llama_cpp,gemma_cpp}. These systems employ optimizations such as quantization and memory-mapped weights to run models with limited resources. Recent studies also focus on improving CPU-aware inference efficiency through sparsity-aware techniques, quantization optimizations, offloading, and library enhancements (e.g., OpenBLAS~\cite{openblas}, oneDNN~\cite{onednn}, XNNPACK~\cite{xnnpack}), some of which are implemented in llama.cpp~\cite{llama_cpp}. \textit{We built our work on top of llama.cpp and come up with new solutions for codesigning resource provisioning and workload scheduling. }

\textbf{Heterogeneous LLM on Commodity Hardware:} Recently, people have gained more interest in deploying or training foundational models under decentralized or heterogeneous commodity setups~\cite{yuan2022decentralized,scao2022bloom}. To account for the heterogeneity in the interconnect bandwidth of commodity setup, Mobius~\cite{Feng2023} introduces a novel pipeline parallelism scheme enabling heterogeneous memory for large-scale model training, while bringing fewer communications than existing systems. 
\textit{While those works mostly focus on model partition and placement, we focus more on the request scheduling and resource provision. Our objective which includes carbon is also fundamentally different from performance-oriented systems. }

\textbf{LLM Serving Offloading:} Recently, memory or compute offloading has gained popularity as a research area due to the capacity limits and high costs of GPU HBM. Many works offload model weights, activations, KV cache, or computations to the CPU for \textit{offline inference} that trade latency for throughput. FlexGen~\cite{flexgen}, HeteGen~\cite{hetegen}, PowerInfer~\cite{song2023powerinfer}, and TwinPilots~\cite{twinpilots} are focused on settings with batch inference of workloads with homogeneous input / output length. 
InstInfer~\cite{instinfer} and DeepSpeed ZeRO-Inference~\cite{aminabadi2022deepspeed} further offloads to SSD, but they did not study the load balance between CPUs and GPUs.
FastDecode~\cite{fastdecode} and NEO~\cite{jiang2024neo} target online serving. NEO proposes asymmetric GPU-CPU pipelining and load-aware scheduling to balance GPU and CPU loads. Specifically for MoE models, Huang et al~\cite{huang_towards_2023} propose to swap experts to CPU memory to reduce the GPU memory consumption. PC-MoE~\cite{kong_serving_2023} proposes to offload experts under dynamic resource constraints with offline profiling and online scheduling. 
SiDA~\cite{Du2023SiDASD} dynamically loads activated experts and offloads inactivated experts according to hash tables.  
\textit{None of the proposals considers the resource provisioning aspect and carbon impact of offloading. Furthermore, naively implementing offloading without considering phases, request lengths, and capacity planning would only cause more embodied and operational carbon due to GPU overprovisioning and large PCIe-induced overhead. }

\textbf{Energy-aware Serving and Model Serving:} A plethora of work in the space of model serving has emerged, focusing on providing SLOs with scheduling, placement, batching and autoscaling solutions. For example, AlpaServe~\cite{li2023alpaserve}, Clockwork~\cite{gujarati2020serving}, Sheperd~\cite{zhang2023shepherd} and Clipper~\cite{crankshaw2017clipper} all are focusing on non-autoregreesive models. However, such general systems often overlook the distinct phases in LLM serving or autoregressive nature of LLMs. 
InFaaS~\cite{romero2021infaas} proposes a way to do model selection and autoscaling. More recently, $\mu$serve~\cite{qiu2024muserve} proposed power-aware LLM serving. 
Splitwise proposed pd disaggregation to optimize for cost, power, and throughput under SLOs ~\cite{patel2023splitwise}. These general systems are not carbon-aware, nor do they consider the role of CPU. \textit{We show in our experiments that carbon-opt is different from energy-opt, and propose co-design solutions for both resource provisioning and runtime scheduling, leaning on the insights of asymmetric carbon cost of CPU and GPUs. }
\section{Conclusion}

Our work introduces a quantitative framework that holistically optimizes AI infrastructure's carbon footprint. EcoServe demonstrates how hardware-software co-design across software provision, scheduling, and hardware provision can reduce total carbon emissions while maintaining performance. Beyond the specific optimizations presented, this work provides a foundation for carbon-aware system design that we hope will inspire future research in sustainable AI infrastructure.

\end{document}